\documentclass[aps,onecolumn,12pt,superscriptaddress]{revtex4}
\usepackage{hyperref}
\usepackage{graphicx,color}
\pdfoutput=1
\usepackage{amsmath}
\usepackage{amssymb}
\usepackage{amsfonts}
\usepackage[font={footnotesize,it}]{caption}

\newcommand{\Tr}{\mathop{\text{Tr}}\nolimits}

\newcommand{\erfc}{\mathop{\text{erfc}}\nolimits}

\newcommand{\be}{\begin{equation}}
\newcommand{\ee}{\end{equation}} 

\begin{document}

\title{Energy-pressure relation for low-dimensional gases}

\author{Francesco Mancarella*}
\affiliation{Nordic Institute for Theoretical Physics (NORDITA), Roslagstullsbacken 23, S-106 91 Stockholm, Sweden}
\affiliation{Department of Theoretical Physics, KTH Royal
Institute of Technology, SE-106 91 Stockholm, Sweden}

\author{Giuseppe Mussardo}
\affiliation{SISSA and INFN, Sezione di Trieste, via Bonomea 265, 
I-34136 Trieste, Italy}
\affiliation{International Centre for Theoretical Physics (ICTP), 
Strada Costiera 11, I-34151, Trieste, Italy}

\author{Andrea Trombettoni}
\affiliation{CNR-IOM DEMOCRITOS Simulation Center, Via Bonomea 265, 
I-34136 Trieste, Italy}
\affiliation{SISSA and INFN, Sezione di Trieste, via Bonomea 265, 
I-34136 Trieste, Italy}

\begin{abstract}
A particularly simple relation of proportionality between internal energy and pressure holds for scale--invariant thermodynamic systems (with Hamiltonians homogeneous functions of the coordinates), including classical and quantum -- Bose and Fermi -- ideal gases. One can quantify the deviation from such a relation by introducing the internal energy shift as the difference between the internal energy of the system and the corresponding value for scale--invariant (including ideal) gases. After discussing some general thermodynamic properties associated with the scale--invariance, we provide criteria for which the internal energy shift density of an imperfect (classical or quantum) gas is a bounded function of temperature. We then study the internal energy shift and deviations from the energy-pressure proportionality in low-dimensional models of gases interpolating between the ideal Bose and the ideal Fermi gases, focusing on the Lieb-Liniger model in $1d$ and on the anyonic gas in $2d$. In $1d$ the internal energy shift is determined from the thermodynamic Bethe ansatz integral equations and an explicit relation for it is given at high temperature. Our results show that the internal energy shift is positive, it vanishes in the two limits of zero and infinite coupling (respectively the ideal Bose and the Tonks-Girardeau gas) and 
it has a maximum at a finite, temperature-depending, value of the coupling. Remarkably, at fixed coupling the energy shift density saturates to a finite value for infinite temperature. In $2d$ we consider systems of Abelian anyons and non-Abelian Chern-Simons particles: as it can be seen also directly from a study of the virial coefficients, in the usually considered hard-core limit the internal 
energy shift vanishes and the energy is just proportional to the pressure, with the proportionality constant being simply the area of the system. Soft-core boundary conditions at coincident points for the two-body wavefunction introduce a length scale, and induce a non-vanishing internal energy shift: the soft-core thermodynamics is considered in the dilute regime for both the families of anyonic models and in that limit we can show that the energy-pressure ratio does not match the area of the system, opposed to what happens for hard-core (and in particular $2d$ Bose and Fermi) ideal anyonic gases.
\end{abstract} 

\maketitle

\begin{small}\textbf{Keywords:}\end{small}\begin{footnotesize} arXiv: 1407.0028;\, Low-dimensional systems;\, Lieb-Liniger model;\, Quantum gases;\, 
Fractional statistics;\, Scale--invariance;\\
 Anyonic thermodynamics;\, Chern-Simons theory;\, Virial expansion \end{footnotesize}\\

\begin{small}\textbf{PACS number:}\end{small} 05.30.-d; 05.30.Pr\\

\begin{small}\textbf{Contact information:}\end{small}\\
\begin{small}* Francesco Mancarella (\textit{corresponding author}):\end{small} \begin{footnotesize}framan@kth.se;\, postal address: NORDITA, Roslagstullsbacken 23, 106 91 Stockholm, Sweden;\, telephone: +46 8 553 784 23 \end{footnotesize}\\
\begin{small}Giuseppe Mussardo:\end{small} \begin{footnotesize}mussardo@sissa.it \end{footnotesize}\\
\begin{small}Andrea Trombettoni:\end{small} \begin{footnotesize}andreatr@sissa.it \end{footnotesize}\\

\newpage

\section{Introduction} \label{section1}

At thermodynamic equilibrium, pure homogeneous fluids (found in regions of the phase diagram hosting a single phase) 
are characterized by an equation of state $f(V,P,T)=0$ relating pressure, volume and temperature. In general, an equation of state 
can be solved with respect to any of the three quantities $V$, $P$, or $T$, thus providing different ways to characterize the equilibrium 
properties of the system: for example, the partial derivatives of the form $V=V(T,P)$ have the physical meaning of thermal expansion coefficient $\alpha_V \equiv (1/V)(\partial V/\partial T)_P$ and isothermal compressibility $\beta_T\equiv - (1/V) (\partial V/\partial P)_T$ \cite{Mayer77,Landau80,Huang87,McQuarrie00}. 

The equations of state for classical and quantum ideal gases are the starting point for understanding the thermodynamics of interacting gases \cite{Mayer77,Landau80,Huang87,McQuarrie00}: in particular, the equation of state for the classical ideal gas is approximately valid for the low-density region of any real gas. In general, the internal energy of an interacting gas is a function of  both temperature and pressure as a result of forces between the molecules. If such forces did not exist, no energy would be required to alter the average intermolecular distance, i.e. no energy would be required to implement volume and pressure changes in a gas at constant temperature. It follows that in the absence of molecular interactions, the internal energy of a gas would depend on its temperature only. These considerations lead to the definition of an ideal gas as the one whose macroscopic behavior is characterized by the two equations:  $PV=Nk_B T$ and $E=E(T)$, where $E$ is the internal energy.

The determination of the deviation of thermodynamic properties of non-ideal gases from the ideal behavior is in general a long-standing problem: a commonly used approach to quantify such a deviation is to define the shift of thermodynamic quantities 
as the difference with respect to the corresponding value of the same quantities in the ideal case. Historically, several 
techniques have been developed in order to encode deviations from the ideal gas law. Equations of states which are cubic in the volume feature a simple formulation together with the ability to represent for instance both liquid and vapor behavior. 
The first cubic equation of states was the Van der Waals equation \cite{Waals1873} 
$$P=k_BT/(v-b)-a/v^2$$ 
($v$ denotes the volume per particle), accounting for attractive intermolecular (or Van der Waals) forces and a finite excluded volume through its positive constants $a$ and $b$ respectively. In the high-temperature regime, the deviations from the ideal equation of state can be expressed in a more general way, called virial expansion, and obtained by expressing the pressure as a power series in the density $\rho$ in the form 
\be \label{virialexpansion}
P = \rho k_B T \left[1 + B_2 (T)\rho + B_3 (T)\rho^2 + \cdots \right],
\ee 
where $B_n (T)$ is the $n$-th virial coefficient \cite{Mayer77,Landau80,Huang87,McQuarrie00}. Many other similar functional forms have been proposed in various contexts for the equation of state of interacting gases, with the virial equations among the first to receive a firm theoretical foundation. In fact, virial expansions can be derived from first principles of statistical mechanics and such a derivation has also the merit to enlighten the physical significance of the various coefficients: the second virial term above written arises on account of interactions between pairs of molecules and, more generally, the $n$-th term depends upon the interactions among $k$-bodies, $k$ ranging from 2 to $n$. 
In this paper we focus on the study of the energy-pressure relation in low-dimensional systems: for ideal gases, the internal energy is simply proportional to the product $PV$ of pressure and volume, with the proportionality constant depending on the dimensionality $d$ of the system. As we discuss in Section \ref{sezionescaleinvariant} this simple relation between energy and pressure holds for any scale--invariant thermodynamic systems, i.e. for systems having a $N$-body Hamiltonian $H_N$ that scales as $H_N \rightarrow \lambda^{-\alpha} H_N$ under a dilatation of $\lambda$-linear scaling factor, and boundary conditions at coincident points on the wave-function $\psi_N(x_1, \cdots, x_N)$ which are true also for any rescaled wave-function $\tilde{\psi}_N(x_1, \cdots, x_N) \equiv \psi_N(\lambda x_1, \cdots, \lambda x_N)$ . The first condition means that $H_N$ is an homogeneous function of the coordinates: ideal classical and quantum gases are particular cases of this class of systems, since their Hamiltonian is scale--invariant with $\alpha=2$.

To quantify deviations from the ideal energy-pressure relation, in the following we introduce the {\em internal energy shift} as the difference between the internal energy of the system and the corresponding value of the scale--invariant (including ideal) gases. Low-dimensional quantum systems provide a natural playground for the study of the internal energy shift,  since in $1d$ and $2d$ systems it is possible to naturally interpolate from the thermodynamic properties of an ideal Bose gas to those of an ideal Fermi gas, and determine how deviations from the ideal gas behavior affect thermodynamic quantities. We will consider the Lieb-Liniger (LL) model in $1d$ and the anyonic gas in $2d$. For these two systems the physical nature of the interpolation between the Bose and Fermi statistics seems to be formally different: 
\begin{itemize}
\item in the LL model (a $1d$ model of interacting bosons), the interpolation between ideal bosonic and fermionic behavior is driven by the increase of the repulsive interaction among the particles. 
\item in $2d$ anionic gases, one can instead explicitly interpolate between the two canonical bosonic and fermionic statistics by tuning the  statistical parameter. 
\end{itemize}
However, the anyonic statistics incorporates the effects of interaction in microscopic bosonic or fermionic systems (statistical transmutation) and, from this point of view, it is again the variation of the underlying microscopic interactions that induces the interpolation between Bose and Fermi ideal gas. 

So, our first paradigmatic example of interpolating behavior between ideal Bose and Fermi gases will be the LL model of 
one-dimensional bosons interacting via a pairwise $\delta$-potential: the equilibrium properties of this model can be exactly solved via Bethe ansatz both at zero \cite{Lieb63} and finite temperature \cite{Yang69}. In the exact solution of this model, a crucial role is played by the coupling $\gamma$, which turns out to be proportional to the strength of the two-body $\delta$-potential: the limit of vanishing 
$\gamma$ corresponds to an ideal $1d$ Bose gas; on the other side, the limit of infinite $\gamma$ corresponds to the Tonks-Girardeau (TG) gas \cite{Tonks36,Girardeau60,Lenard64}, having (local) expectation values and thermodynamic quantities of a $1d$ ideal Fermi gas 
\cite{Yurovsky08,Pethick08,Bouchoule09,Cazalilla11}. Two features makes the LL model attractive for the purposes of studying the 
internal energy: first, its integrability \cite{Korepin93,Samaj13}, crucial for getting non-perturbative exact results all along the crossover from weak to strong coupling regimes; second, its experimental realization by means of ultracold atom set-ups \cite{Olshanii98,Yurovsky08,Pethick08,Bouchoule09,Cazalilla11}, where bosons are confined within $1d$ atom waveguides which freeze almost all transverse degrees of freedom 
\cite{Kinoshita04,Paredes04,Vandruten08}. The coupling strength of the LL system can be tuned through the Feshbach resonance mechanism \cite{Astrakharchik05}.

Our second paradigmatic example will be the $2d$ ideal anyonic gases in which we will study the energy-pressure 
relation in the interpolation between $2d$ Bose and Fermi gases induced by the pure statistical Aharonov-Bohm interactions.  
We will consider Abelian and non-Abelian Chern-Simons particle systems, and both models admit a soft-core generalization that 
can be understood as the result of an additional contact interaction besides the pure statistical one. As it is well known, 
quantum two-dimensional systems of indistinguishable particles have the peculiarity of admitting generalized braiding statistics, 
because of the non-trivial topological structure of braiding transformations defined over the space-time ambient manifold. Ordinary bosonic and fermionic quantum statistics in $2d$ admit the generalization represented by Abelian anyons, where an elementary braiding operation is encoded in terms of a multiplicative phase factor acting on the 
multi-anyonic scalar wavefunction \cite{Wilczek90,Lerda92,Forte92,Khare05,Nayak08}. 
A different generalization of the standard quantum statistics is represented 
by non-Abelian anyons, described by a multi-component many-body wavefunction and corresponding to higher-dimensional representations of the braid group: non-Abelian anyons generalize the parastatistics, exactly in the same manner 
in which Abelian anyons generalize Bose and Fermi statistics. 

Thermodynamic properties of ideal Abelian anyonic gas (assuming hard-core boundary conditions for the wavefunction 
at coincident points) were studied in the low-density regime \cite{Arovas85}: the exact expression therein obtained for the second virial coefficient is periodic and non-analytic as a function of the statistical parameter. Different approaches have been subsequently used in order to approximate the values of a few higher virial coefficients, including the semiclassical approximation \cite{Bhaduri91} and Monte Carlo computations \cite{Myrheim93} (for more references see \cite{Dasnieres92,Khare05}). The thermodynamics of a system of free non-Abelian anyons appears as a harder task and, so far, only results about the second virial coefficient are available 
\cite{Lo93,Lee95,Hagen96,Mancarella13}. In Section \ref{otherthermod_2d} we also study the shift of the internal energy of soft-core anyonic gases: a family of models for "colliding" anyons (featuring generalized soft-core boundary conditions) can be introduced as the set of well-defined self-adjoint extensions of the Schr\"odinger anyonic Hamiltonian. The mathematical arguments underlying the possibility of such a generalization were discussed in \cite{Bourdeau92}, and the second virial coefficient 
of soft-core Abelian anyons was studied in \cite{Manuel91,Giacconi96,Kim98}. 
The corresponding self-adjoint extensions for the non-Abelian anyonic theory have been as well discussed 
\cite{Bak94,Amelino95,Lee97,Mancarellainterparticlepotential}. The model of soft-core anyons is here considered as an explicit example of scaling symmetry breaking due to the presence of an intrinsic length scale. 

Among all thermodynamic properties of ideal classical and quantum gases, the linear relation between internal energy and pressure is particularly simple, and in this paper we study how it is affected by the various interactions represented by the low-dimensional models above mentioned. For extensive computer simulations of energy-pressure relation in 3D classical systems of interacting particles, see \cite{Bailey08,Bailey08second,Schroder09,Gnan09,Schroder11,Dyre13} (and \cite{Schroder14} for the definition of "Roskilde systems").

The paper is organized as follows: in Section \ref{sezionescaleinvariant} we show that a simple relation of proportionality between internal energy and pressure holds for scale--invariant thermodynamic systems, including classical and quantum (Bose and Fermi) ideal gases, and we discuss some simple consequences of scale--invariance in generic dimensionality, 
including some useful properties of isoentropic transformations. In Section \ref{sezionegasimperfetti} we set criteria under which 
the internal energy shift per particle of an imperfect gas at fixed density saturates towards a finite value as the temperature becomes very large. These criteria are expressed in terms of the second virial coefficient and by distinguishing the different dimensionalities.  
Section \ref{themodel} is devoted to define the models we are going to study in next Sections: the LL models and the different 
anyonic models. The internal energy shift of the LL model is studied in Section \ref{otherthermod} by using thermodynamic Bethe 
ansatz integral equations: the comparison with the $1d$ hard-core bosons is also discussed. Section \ref{otherthermod_2d} 
deals with the internal energy shift of anyonic gases, and we present results for both the hard- and the soft- core anyonic gases. 
Our conclusions are drawn in Section \ref{conclusions}, while more technical material is presented in the Appendices.

\section{Scale--invariant systems}\label{sezionescaleinvariant}
A proportionality between internal energy $E$ and pressure $P$ holds for any scale--invariant thermodynamic system. 
Indeed, let us consider a (classical or quantum) system of $N$ particles in a volume $V$ with Hamiltonian $H_N(V)$. It is intended 
that in this Section and the next, we denote by $V$ the length $L$ in $1d$ and the area $A$ in $2d$. We define a classical system 
to be scale--invariant when the Hamiltonian transforms as 
\begin{equation}
H_N\rightarrow \lambda^{-\alpha} H_N
\label{scalingHamiltonianN}
\end{equation} 
under a dilatation of a $\lambda$-linear scaling factor such that the coordinate $x$ of the particles transforms 
as $x \rightarrow \lambda x$ (the momentum $p$ transforms correspondingly as $p \rightarrow \lambda^{-1} p$): therefore 
the Hamiltonian is an homogeneous function of its spatial coordinates. 
For quantum systems, we define them to be scale--invariant if they fulfill condition (\ref{scalingHamiltonianN}) and respect at the same time scale--invariant boundary conditions for the $N$-body wave-function $\psi_N$ at contact points, i.e conditions on the wave-function $\psi_N(x_1, \cdots, x_N)$ which are true also for any rescaled wave-function 
$\tilde{\psi}_N(x_1, \cdots, x_N)) \equiv \psi_N(\lambda x_1, \cdots, \lambda x_N)$, where $\lambda\neq 0$ is a real constant. 
A typical example of scale--invariant boundary conditions for the $N$-body wave-function at contact points is given by the hard-core condition.

In the canonical ensemble the pressure $P$ and the internal energy $E$ are defined as
\be \begin{array}{l} 
P=\frac{1}{\beta} \frac{\partial}{\partial V} \log{Z(N,V,\beta)}\,,\\
E=-\frac{\partial}{\partial \beta} \log{Z(N,V,\beta)}\,,\\
\end{array}
\label{relaztermodinamicheperinvarianza}
\ee
where as usual $\beta=1/k_B T$ and $Z$ is the partition function:
\be
Z(N,V,\beta)= \Tr e^{-\beta H_N(V)}\,.
\label{part}
\ee
For any $d$-dimensional scale--invariant system of volume $V$, the map 
\be
(V,\beta) \rightarrow (\lambda^d V,  \lambda^\alpha \beta)
\label{map}
\ee 
leaves $\log Z$ invariant in the thermodynamic limit (since $\beta H_N \rightarrow \beta H_N$). 
With the notation $\lambda-1=\epsilon \ll 1$, we are led to
\be 
0\,=\,(\delta\log Z)\vert_{(\delta V,\delta\beta)\,=\,(\epsilon  d V ,\epsilon \alpha\beta)}\,=\,\epsilon\left(V \,d\,\frac{\partial}{\partial V}\log Z+\alpha\,\beta \frac{\partial}{\partial \beta}\log Z\right)\;,
\label{variazionegeneralepartizione}
\ee 
whence relations (\ref{relaztermodinamicheperinvarianza}) imply
\be
E=\frac{d}{\alpha}\,P\,V\,. 
\label{relazionefondamentale}
\ee
Notice that Eq.\,(\ref{relazionefondamentale}) is valid both for classical and quantum scale--invariant systems, and follows from the invariance of the partition function under map (\ref{map}): therefore the scale--invariance of boundary conditions at contact points is required in the quantum case. So, for instance, in Section \ref{otherthermod_2d}, we will show that, for the $2d$ ideal anyonic gas, Eq.(\ref{relazionefondamentale}) only holds in the case of hard-core boundary conditions while it is violated in the soft-core case. From Section \ref{themodel} onwards, we study some low-dimensional quantum systems, since we are primarily interested in interpolating between the two ordinary quantum statistics. 

From the considerations above, it follows that any quadratic scale--invariant Hamiltonian fulfills the scaling property 
$H_N\rightarrow \lambda^{-2} H_N$ under a dilatation of a $\lambda$-linear scaling factor and therefore 
enjoys the property 
\be
E=\frac{d}{2}\,P\,V\,.
\label{usual}
\ee 
Few examples of systems for which property (\ref{usual}) is known are the following: 
\begin{itemize}
\item[{\em i)}] 
$d$-dimensional ideal classical and quantum (Bose and Fermi) gas have quadratic dispersion relations, and they all obey the well known 
relation $E=(d/2)P\,V$, as it can be also deduced from the virial theorem \cite{Uhlenbeck32}; 
\item[{\em ii)}] 
as reviewed in Section \ref{otherthermod_2d}, the $1d$ LL Bose gas has total internal energy $E=PL/2$ (since $d=1$ and $\alpha=2$) 
for any temperature in both its scale--invariant limit regimes: 
the non-interacting limit 
($\gamma \rightarrow 0$) and the fermion-like Tonks-Girardeau limit ($\gamma\rightarrow \infty$), which correspond respectively to 
the zero and infinite coupling associated with the $\delta$-like contact interactions; 
\item[{\em iii)}] for the $3d$ Fermi gas at the unitary limit the relation $E=(3/2)PV$ holds as well \cite{Ho04} (see the discussion in \cite{Castin12}).
\end{itemize}
As discussed in Section \ref{otherthermod_2d}, also $2d$ hard-core ideal anyonic gases obey Eq.\,(\ref{relazionefondamentale}) for general values of the statistics parameters.

We derive now some scaling properties for scale--invariant $d$-dimensional systems undergoing adiabatic reversible thermodynamic processes (as above, the argument is carried out in the quantum case for the sake of generality). Let us consider the scale--invariant thermodynamic system confined in a region subjected to a quasi-static scaling transformation of the volume and 
the temperature $(V,T)\rightarrow (\lambda^d\, V,\, \lambda^{-\alpha}\, T)$, under which the ratios $E_i/k_B T$ are left invariant (same proof of (\ref{formulaspettro})), 
as long as the $N$-particle Hamiltonian $H_N$ gains a $\lambda^{-\alpha}$ factor under a $\lambda$-factor scaling of its spatial coordinates. 
The total entropy $S$ of the system remains invariant under such a process: indeed the energy
\be
E\equiv \frac{\sum_i e^{-E_i/k_B T}\,E_i}{\sum_i e^{-E_i/k_B T}}
\ee
scales proportionally to $\lambda^{-\alpha}$ (because of the transformation $(T,\{E_i\})\rightarrow (\lambda^{-\alpha} T,\{\lambda^{-\alpha} E_i\})$ 
of temperature and energy levels), exactly as required for any isoentropic process fulfilling relation (\ref{relazionefondamentale}). 
This last statement results from $E=\frac{d}{\alpha}PV$ and $P=-\partial_V E(N,S,V)$, which give $(\alpha/d)E_{isoentr}/V=-(dE_{isoentr}/dV)$ and therefore:
\be
E\,V^{\alpha/d}=const\;,
\ee
i.e.	 $E\propto\lambda^{-\alpha}$, along the series of equilibrium state of a given isoentropic process.
We conclude that adiabatic reversible expansions and compressions (as well as arbitrary isoentropic processes followed by thermal relaxation) of 
scale--invariant systems are characterized by the following transformations for internal energy and temperature:
\be 
E \propto V^{-\alpha/d}\,;\quad T\propto V^{-\alpha/d}\,.
\label{trasformazioneisoentropica}
\ee
It is worth to point out three immediate consequences of (\ref{trasformazioneisoentropica}):
\begin{itemize}
\item as ideal gases, scale--invariant systems undergoing an isoentropic process comply with the invariance of $PV^{\tilde{\gamma}}$, 
where $\tilde{\gamma}\equiv 1+\alpha/d$.
\item isoentropic transformations of scale--invariant systems let the dilution parameter $x \equiv \rho \lambda_T^d$ invariant, 
by taking into account a generalized definition of the thermal wavelength depending on the dispersion relation and the dimensionality \cite{Yan00}.
\item the internal energy associated with equilibrium states of an isoentropic process is proportional to the temperature:
\be 
E=y \times Nk_BT\,,
\label{proporzenergiatemperatura}
\ee
where $y$ remains constant along the isoentropic curve. Notice that the factor $y$ would depend solely on the dilution parameter $x=\rho\,\lambda_T^d$ for a given system if Eq. (\ref{proporzenergiatemperatura}) is considered over the entire phase diagram.
\end{itemize}

It is worth to point out that for scale--invariant Hamiltonian systems, the dependence of virial coefficients upon the temperature is very simple, i.e. 
\be
B_k(T) \propto T^{-\frac{d}{\alpha}(k-1)}
\label{relation_B_T} \,\,\,; 
\ee
in fact, the parameters of an homogeneous Hamiltonian (\ref{scalingHamiltonianN}) define only a set of independent scales $\{a_n\}$ having dimensions energy $\times$ (length)$^\alpha$. By definition (\ref{virialexpansion}), the corresponding virial coefficients $B_k(T)$ have dimensionality $d(k-1)$, therefore their temperature-dependence has to be of the form (\ref{relation_B_T}). Furthermore, for several scale--invariant quantum systems (such as Fermi and Bose gas, unitary Fermi gases and a large variety of systems definable in terms of vector interactions, e.g. Abelian anyons and various kind of non-Abelian anyons) the thermal length $\lambda_T=h/\sqrt{2\pi m k_BT}$ is the only inherent length scale defined in terms of their parameters, and as a consequence Eq. (\ref{relation_B_T}) takes for them the special form $B_k(T)\propto \lambda_T^{d(k-1)}$. As it will be discussed in Section \ref{otherthermod_2d}, the fact that the $B_k(T)$ respect Eq.\,(\ref{relation_B_T}) 
implies the validity of the relation (\ref{relazionefondamentale}) at all orders of the virial expansion (within its radius of convergence).

We conclude this Section by showing that it is also possible to deduce a property of the internal pressure for scale--invariant systems. 
The internal pressure $\pi_T$ is defined in general as the volume derivative of the internal energy in isothermal processes \cite{Atkins10}:
\be \label{defintpressure}
\pi_T  = \left ( \frac{\partial U}{\partial V} \right )_T\,;   
\ee
the internal pressure is a measure of attractiveness for molecular interactions and is related 
to the (thermodynamic) pressure $P$ by the expression
\be \label{thermodeqstate}
\pi_T = T \left ( \frac{\partial P}{\partial T} \right )_V - P\;.
\ee
Eq. (\ref{thermodeqstate}) is usually referred to as the {\em thermodynamic equation of state}, because 
it expresses the internal pressure just in terms of fundamental thermodynamic parameters $P,V,T$.
For general scale--invariant systems, plugging (\ref{relazionefondamentale}) and (\ref{relation_B_T}) into the 
definition (\ref{defintpressure}) [or equivalently Eq. (\ref{relation_B_T}) 
into Eq. (\ref{thermodeqstate})] leads to the following result in the dilute regime:
\be \label{intpressscaleinv}
 \pi_T \sim -\frac{d}{\alpha}\,\delta P,\quad\quad \delta P\equiv P-P_{ideal}\;, 
\ee
where $P_{ideal}$ is the pressure of the ideal Boltzmann gas having the same $V,T$.

Relation (\ref{intpressscaleinv}) does not hold, in general, if scale--invariance is violated. For scale--invariant 
systems with dimensionality equal to the dispersion-relation exponent (such as $2d$ Bose, $2d$ Fermi and hard-core Abelian and non-Abelian anyonic gases for quadratic dispersion), Eq.(\ref{intpressscaleinv}) reads $\pi_T \sim -\delta P$. In such systems the positive(/negative) internal pressure (\ref{defintpressure}) can be exactly regarded in the dilute limit as the interaction contribution 
acting in favor to(/against) the external pressure $P$ and counterbalancing the thermal contribution $P_{Boltzmann}$.

\section{Energy-pressure relation for imperfect gases at high temperature}\label{sezionegasimperfetti}
Hereafter we denote by $\rho$ the number density, by $E_{res}$ the 
internal energy shift and by $e_{res}(\rho,T)$ the internal energy shift density 
(per particle) of a generic classical or quantum imperfect gas, defined as 
\be E \equiv \frac{d}{\alpha}PV+E_{res} \equiv \frac{d}{\alpha}PV+Ne_{res}\;.
\label{defresidualenergy}
\ee
The internal energy shift represents a measure of the deviation from the relation (\ref{relazionefondamentale}) derived in Section \ref{sezionescaleinvariant} for scale--invariant systems whose Hamiltonians are homogeneous functions of the coordinates. 

In many textbooks, deviations from ideal gas behavior are quantified by introducing the so-called 
{\em departure functions} (or {\em residual} thermodynamic quantities) (see, for instance, \cite{Smith00}). 
Such departure functions are obtained by taking the difference between the considered quantity 
and the corresponding value for the ideal gas, when two among the $P$, $V$ and $T$ parameters 
are kept fixed, typically $P$ and $T$ \cite{Smith00}. The quantity defined in (\ref{defresidualenergy}) is a 
departure internal energy, but with $V$ and $P$ fixed: however, the departure (or residual) internal energy conventionally 
defined fixing $P$ and $T$ [i.e. defined as $E-(d/2) N k_ B T$] is not zero 
for general scale--invariant systems, while with our definition (\ref{defresidualenergy}) 
of the internal energy shift, the latter vanishes for all scale--invariant systems. (It can be immediately checked that the only scale--invariant system whose conventionally defined departure internal energy vanishes is the ideal gas). An example 
of a system which is scale--invariant but with non-vanishing (conventionally defined) departure internal energy is 
the hard-core anyonic gas, as discussed in Section \ref{otherthermod_2d}: on the contrary, the one defined in (\ref{defresidualenergy}) can be considered as the correct residual quantity measuring deviations from scale--invariance. To avoid possible misunderstandings, we decided to refer to $E_{res}$ [defined in (\ref{defresidualenergy})] 
as the {\em internal energy shift} rather than {\em departure internal energy}. 

In the low-density regime, the thermodynamic quantities can be associated with the virial coefficients $\{ B_n(T) \}$ of the equation of state $P=P(\rho,T)$. 
Following statistical mechanics textbooks \cite{Mayer77,Landau80,Huang87,McQuarrie00}, 
in the $d$-dimensional case the following virial expansions for the pressure $P$, the Helmholtz free energy $A_H$, 
the Gibbs free energy $G$, the entropy $S$, the internal energy $E$ 
and the enthalpy $H$ are obtained:
\be \begin{array}{l} \text{Pressure}: \,\,\,\,\,\,\frac{PV}{Nk_BT}=1+\sum\limits_{k \geq 1} B_{k+1}\, \rho^k\,\,\,;\\
\text{Helmholtz free energy}:\,\,\,\,\,\, \frac{A_H}{Nk_BT}=\frac{d}{2}-\frac{S_{ideal}}{Nk_B}+\sum\limits_{k \geq 1} \frac{1}{k}\, B_{k+1}\, \rho^k\,\,\,;\\ 
\text{Gibbs free energy}:\,\,\,\,\,\, \frac{G}{Nk_BT}=\frac{d}{2}+1-\frac{S_{ideal}}{Nk_B}+\sum\limits_{k \geq 1} \frac{k+1}{k}\, B_{k+1}\, \rho^k\,\,\,;\\ 
\text{Entropy}: \,\,\,\,\,\, \frac{S}{Nk_B}=\frac{S_{ideal}}{Nk_B}-\sum\limits_{k \geq 1} \frac{1}{k}\, \frac{\partial}{\partial T}\left(T\, B_{k+1}\right)\, \rho^k\,\,\,; \\ 
\text{Internal energy}: \,\,\,\,\,\, \frac{E}{Nk_BT}=\frac{d}{2}-T\,\sum\limits_{k \geq 1} \frac{1}{k}\, \frac{\partial B_{k+1}}{\partial T}\, \rho^k\,\,\,;\\ 
\text{Enthalpy}: \,\,\,\,\,\, \frac{H}{Nk_BT}=\frac{d}{2}+1+\sum\limits_{k \geq 1} \left( B_{k+1}-\frac{1}{k}\,T\,\frac{\partial B_{k+1}}{\partial T}\right)\, \rho^k\,\,\,.
\end{array}
\label{zeroset}
\ee
Below we state the necessary conditions (proven in Appendix \ref{condizioneenergiaresidua}) under which the energy shift of a (classical or quantum) gas remains bounded in the limit of high temperatures, i.e.
\be \label{boundedness}
\lim\limits_{T\rightarrow \infty}\vert e_{res}(\rho,T)\vert<\infty\;.
\ee
For simplicity, hereafter, we limit ourselves to the case of quadratic dispersion relation $\alpha=2$, for which  
such conditions are (with $c_1$ and $c_2$ real coefficients):
\begin{itemize}
\item For $d=1$: \be \label{criterio1}
B_2(\beta)=c_1\sqrt{\beta}+c_2 \beta + o(\beta), \quad\quad \text{ and in this case } \lim\limits_{T\rightarrow \infty}e_{res}(\rho,T)=\frac{c_2}{2}\rho\;.
\ee
In Section \ref{otherthermod} the explicit expression of $B_2$ for the LL model as a function 
of the coupling constant $\gamma$ \cite{Yang70} is reported: for finite $\gamma$, it is in general $c_2 \neq 0$, so that $e_{res}$ is bounded.
  
\item For $d=2$: \be \label{criterio3}
B_2(\beta)=c_3\,\beta \log \beta +o(\beta \log \beta), \quad\quad \text{ and in this case } \lim\limits_{T\rightarrow \infty}e_{res}(\rho,T)=c_3 \, \rho\,.
\ee
In Section \ref{otherthermod_2d} the 2D anyonic gas is studied, and shown to have vanishing internal energy shift in the high-temperature limit. This is in agreement with (\ref{criterio3}) because, referring as $\alpha$ to the statistical parameter, $B_2^{h.c.}(\alpha, \beta)= c_1(\alpha)\,\beta$ \cite{Arovas85},\quad  $B_2^{s.c.}(\alpha, \beta)=c_1'(\alpha)\,\beta+c_2(\alpha)\, \beta^{1+\vert\alpha\vert}+o(\beta^{1+\vert\alpha\vert})$ \cite{Kim98, Mancarella13}, where $c_1,c_1',c_2$ are suitable functions, hence both are subleading w.r.t. $\beta \log \beta$ in the $\beta\rightarrow 0$ limit.

\item For $d > 2$ : 
\be \label{criterio2}
\quad\quad\quad\quad\quad\quad\;\; B_2(\beta)=c_2 \beta + o(\beta), \quad\quad \text{ and in this case } \lim\limits_{T\rightarrow \infty}e_{res}(\rho,T)=
\left(1-\frac{d}{2}\right) \, c_2 \, \rho\,.
\ee

\end{itemize}



\section{The Models}\label{themodel}
In this Section we recall the main properties of the LL and anyonic models studied in the next Sections: in Subsection \ref{sezioneLiebLiniger} we introduce the $1d$ Lieb-Liniger model, in Subsection \ref{abeliananyons} we outline the main thermodynamic properties of an ideal gas of Abelian anyons (and its soft-core generalization), while in Subsection \ref{nonabeliananyons} we briefly introduce the system of Non-Abelian Chern Simons (NACS) particles, i.e. a model of non-Abelian anyons.

\subsection{Lieb-Liniger model}\label{sezioneLiebLiniger}
The LL Bose gas is described by an Hamiltonian for $N$ non-relativistic bosons of mass $m$
in one dimension interacting via a pairwise $\delta$-potential \cite{Lieb63} having the form
\be
\label{HAM_LL}
H=-\frac{\hbar^2}{2m}\sum_{i=1}^N\frac{\partial^2}{\partial x_i^2}+2\lambda\,
\sum_{i<j}\delta(x_i-x_j)\,, 
\ee
where $\lambda$ is the strength of the $\delta$-like repulsion (we consider here 
only positive or vanishing values of $\lambda$: $\lambda \ge 0$). 
The effective coupling constant of the LL model is given by the
dimensionless quantity
\begin{equation}
\gamma=\frac{2m\lambda}{\hbar^2 \rho}\,, 
\label{eq:gamma}
\end{equation}
where $\rho=N/L$ is the density of the gas. We also use the notation
\begin{equation}
c=\frac{2m\lambda}{\hbar^2}\,, 
\label{eq:c}
\end{equation}
so that $\gamma=c/\rho$. The limit $\gamma \ll 1$ corresponds to the weak coupling limit: in
this regime the Bogoliubov approximation gives a good estimate of the ground-state energy of the system \cite{Lieb63}. For large $\gamma$ one approaches instead the Tonks--Girardeau limit \cite{Girardeau60}. 

In the LL model temperatures are usually expressed in units of the quantum degeneracy temperature $T_D$ as
$$\tau=\frac{T}{T_D}\,,$$
where
\begin{equation}
k_B T_{D}=\frac{\hbar^2 \rho^2}{2m}\,.
\end{equation}

The thermodynamic Bethe ansatz integral equations relate at temperature $T$ the pseudo-energies $\varepsilon(k)$ 
to density $f(k)$ of the occupied levels \cite{Yang69,Korepin93,Samaj13}. One has the following set of coupled equations
\begin{subequations}
\begin{align}
&\varepsilon(k)=-\tilde{\mu}+\frac{\hbar^2 k^2}{2m}-k_B T \int_{-\infty}^{\infty} 
\frac{c/\pi}{(k-k')^2+c^2} \log{\left( 1+e^{-\varepsilon(k')/k_B T} \right)}\,dk'\\
&\rho=\int_{-\infty}^{\infty} f(k)\,dk\,,\\
&f(k) \left( 1+e^{\varepsilon(k)/k_B T} \right) = \frac{1}{2\pi} + \int_{-\infty}^{\infty} 
\frac{c/\pi}{(k-k')^2+c^2} f(k') \, dk'\,,\label{eq:YYTBA1c}
\end{align}
\label{eq:YYTBA1}
\end{subequations}
where $\tilde{\mu}$ is the chemical potential. At $T=0$ the energy level density gets a compact support, so that Eq.\,(\ref{eq:YYTBA1c}) becomes 
\be
f(k) = \frac{1}{2\pi} + \int_{-K}^{K} 
\frac{c/\pi}{(k-k')^2+c^2} f(k') \, dk'\,,
\label{eq:YYTBA1_T0}
\end{equation}
where the boundary value $K$ has to be determined from the condition
\be
\rho=\int_{-K}^{K} f(k)\,dk\,.
\label{eq:norm_T0}
\end{equation}

If one measures energies in units of $k_BT_D$ and wave-vectors in units of $\rho$, by defining 
the scaled wave-vector $\mathcal{K} \equiv k/\rho$, the scaled pseudo-energies 
$E(\mathcal{K}) \equiv \varepsilon(k)/k_BT_D$ and the scaled potential 
$\mu \equiv \tilde{\mu}/k_B T_D$, Eqs. (\ref{eq:YYTBA1}) read
\begin{subequations}
\begin{align}
&E(\mathcal{K})=-\mu+\mathcal{K}^2- \tau \int_{-\infty}^{\infty} 
\frac{\gamma/\pi}{(\mathcal{K}-\mathcal{K}')^2+\gamma^2} \log{\left( 1+
e^{-E(\mathcal{K}')/\tau} \right)}\,d\mathcal{K}'\\
&1=\int_{-\infty}^{\infty} f(\mathcal{K})\,d\mathcal{K}\,,\\
&f(\mathcal{K}) \left( 1+e^{E(\mathcal{K})/\tau} \right) = \frac{1}{2\pi} + \int_{-\infty}^{\infty} 
\frac{\gamma/\pi}{(\mathcal{K}-\mathcal{K}')^2+\gamma^2} f(\mathcal{K}') \, d\mathcal{K}'\,.
\end{align}
\label{eq:YYTBA1_scal}
\end{subequations}
One sees that scaled quantities depends only on $\gamma$ and $\tau$.
 
Once the TBA integral equations (\ref{eq:YYTBA1_scal}) are solved, thermodynamic quantities as the free energy can be computed. In Section \ref{otherthermod} we report both the expressions of internal energy and pressure, and we study the internal energy shift.

\subsection{Abelian Anyons}\label{abeliananyons} 
The dynamics of a systems of $N$ identical Abelian anyons is expressed by 
\cite{Khare05} 
\be \label{HNanioniabeliani}
H_N=\frac{1}{2M}\sum_{i=1}^{N}(\vec{p}_i-\alpha\,\vec{a}_i)^2\,,
\ee
where $$\vec{a}_i=\hbar \,\sum_{j\neq i}\nabla_i \theta_{ij}\;,$$ with $\theta_{ij}$ the relative angle between the particles $i$ and $j$. 
The study of the thermodynamics for a system of identical Abelian anyons has been developed starting with \cite{Arovas85}, 
in which the exact quantum expression for the second virial coefficient has been derived: 
\begin{equation} 
B_2^{h.c.}(2j+\delta,T)=-\frac{1}{4}\lambda_T^2+\vert\delta\vert\lambda_T^2-\frac{1}{2} \delta^2\lambda_T^2\,.\label{BdiArovas} 
\end{equation} 
Eq. (\ref{BdiArovas}) holds provided that hard-core wavefunction boundary conditions are assumed, i.e. $\lim\limits_{x_i\rightarrow x_j} \psi_N(x_1,\cdots, x_N)=0$ for any $1\leq i<j \leq N$, being $\psi_N$ the $N$-body wavefunction in the bosonic gauge \cite{Khare05}.  

In (\ref{BdiArovas}) 
$\alpha=2j+\delta$, where $\alpha$ 
represents the statistical parameter of anyons \cite{Khare05}, 
$j$ is an integer 
and $\vert\delta\vert\leq 1$. We remind that $\alpha=1$ and $\alpha=0$ 
correspond respectively to free $2d$ spin-less fermions and bosons, and that 
$\lambda_T$ is the thermal wavelength defined as 
\be
\lambda_T=\sqrt{\frac{2 \pi \hbar^2}{M k_B T}}\,.
\label{de_B}
\ee
 
The virial expansion is expressed in powers of the number density $\rho$; in the dilute regime, the second virial coefficient gives the leading contribution to the deviation of the energy-pressure relation from the non-interacting case, as a result of rewriting the grand canonical partition function as a cluster expansion \cite{Mayer77,Huang87}.
About the higher virial coefficients of the ideal anyonic gas, only numerical approximations of the first few ones are available so far  \cite{Khare05}, and they are limited to the hard-core case.

The {\em relative} two-body Hamiltonian for a free system of anyons with statistical parameter $\alpha$, 
written in the bosonic description, is of the form \cite{Khare05}  
\begin{equation} 
H_{rel}=\frac{1}{M}(\vec{p}-\alpha \vec{A})^2\,, 
\end{equation} 
where $\vec{A}=(A^1,A^2)$ and $A^i\equiv\frac{\hbar\, \epsilon^{ij}x^j}{r^2}$ ($i=1,2$ and $\epsilon^{ij}$ is the completely antisymmetric tensor). Without any loss of generality, the statistical parameter $\alpha$ can be chosen as $\alpha \in [-1,1]$ \cite{Khare05}. 
By relaxing the regularity condition on the wavefunctions at contact points, 
it is possible to obtain the one-parameter family of soft-core boundary 
conditions (\ref{scboundaryconditions}), according to the method of self-adjoint extensions \cite{Albeverio88}. The $s$-wave solutions of the radial Schr\"odinger equation correspond to a one-parameter family of boundary conditions \cite{Manuel91,Kim98}: 
\be \label{scboundaryconditions}
\lim\limits_{r\rightarrow 0} \left\{r^{\vert\alpha\vert} R_0(r)-\frac{\sigma}{\kappa^{2\vert\alpha\vert}} \frac{\Gamma(1+\vert\alpha\vert)}{\Gamma(1-\vert\alpha\vert)} \frac{d}{d (r^{2\vert\alpha\vert})}\left[r^{\vert\alpha\vert}R_0(r)\right] \right\}\;,
\ee
and correspondingly read as 
\begin{equation} \label{sol} 
R_0(r) = \text{const} \cdot \, \left[ 
        J_{|\alpha|}(kr) + \sigma \left(\frac{k}{\kappa}\right)^{2|\alpha|}J_{-|\alpha|}(kr) 
        \right]\,, 
\end{equation}
where $\sigma=\pm1$ and $\kappa$ is a momentum scale introduced by the boundary condition. 

We refer to 
\begin{equation}\label{hard-core-par}
\varepsilon \equiv \frac{\beta \kappa^2}{M}
\end{equation} 
as the hard-core parameter of the gas. 
If $\sigma=-1$, 
in addition to the solution (\ref{sol}), there is a bound state with energy $E_B=-\varepsilon k_B T=-\kappa^2 / M$ and wavefunction  
\begin{equation} 
R_0(r)= \text{const} \cdot \, K_{|\alpha|}(\kappa r)\,.\label{boundstate} 
\end{equation}  
The second virial coefficient for Abelian anyons in this general case has been computed through different approaches in \cite{Giacconi96,Moroz96,Kim98}, and is given by  
\begin{equation} 
\label{explicitintegralform} 
B_2^{s.c.}(T)=B_2^{h.c.}(T)- 2 \lambda_T^2 
\left\{ 
e^\varepsilon \theta(-\sigma)  
+\frac{\alpha\sigma}{\pi} \sin{\pi\alpha} 
\int_0^\infty \frac{dt\, e^{-\varepsilon t}\, t^{|\alpha|-1}}{1+2\sigma\cos{\pi\alpha}\;t^{|\alpha|}+t^{2|\alpha|}} \right\}\;,
\end{equation} 
where $\theta(x)$ is the Heaviside step function and $B_2^{h.c.}$ is the hard-core result (\ref{BdiArovas}). For $\sigma=+1$, $\varepsilon\rightarrow \infty$ one retrieves the hard-core case ($\psi(0)=0$).
The hard-core limit corresponds to scale--invariance \cite{Jackiw90,Bourdeau92,Kim95,Kim97}, 
while for any other boundary condition (i.e., self-adjoint extension of the Hamiltonian) the characteristic 
scale can be put in relation with the hard-core parameter defined in (\ref{hard-core-par}). 
  
\subsection{Non-Abelian Anyons}\label{nonabeliananyons} 
The $SU(2)$ non-Abelian Chern-Simons (NACS) spin-less particles are point-like sources mutually interacting 
via a topological non-Abelian Aharonov-Bohm effect \cite{Dunne99}. These particles carry non-Abelian charges and non-Abelian magnetic fluxes, so that they acquire fractional spins and obey braid statistics as non-Abelian anyons. 

Details on NACS statistical mechanics \cite{Guadagnini90,Verlinde91,Lee93,Kim94,Bak94} are given in Appendix \ref{nacsqsm} for general soft-core boundary conditions \cite{Amelino95,Lee97}. For non-Abelian anyons, the independence on the statistics of the virial coefficients in a strong magnetic field has been established in \cite{Polychronakos00} while the theory of non-relativistic matter with non-Abelian Chern-Simons gauge interaction in (2+1) dimensions was studied in \cite{Cappelli95}. The $N$-body Hamiltonian for ideal  non-Abelian Chern-Simons quantum particles can be written as \cite{Bak94} 
\begin{equation} 
{H}_N=-\sum_{\alpha=1}^{N} \frac{1}{M_\alpha}\left(\nabla_{\bar 
z_\alpha}\nabla_{z_\alpha}  +\nabla_{z_\alpha}\nabla_{\bar 
z_\alpha}\right)\,,
\label{hamiltonianadelmodello}
\end{equation}
where $M_\alpha$ is the mass of the $\alpha$-th particles, 
$\nabla_{\bar z_\alpha}=\frac{\partial}{\partial \bar z_\alpha}$ and 
\begin{equation}
\nabla_{z_\alpha}=\frac{\partial}{\partial z_\alpha}  +\frac{1}{2\pi 
\kappa} \sum_{\beta\not=\alpha} \hat Q^a_\alpha \hat Q^a_\beta \frac{1}{ 
z_\alpha -z_\beta}\,\,\,.
\end{equation}
In Eq.\,(\ref{hamiltonianadelmodello}) $\alpha = 1, \dots, N$ labels the particles, $(x_\alpha, y_\alpha)=(z_\alpha+\bar z_\alpha, 
-i(z_\alpha-\bar z_\alpha))/2$ are their spatial coordinates, and $\hat Q^a$'s are the isovector operators in a representation of isospin $l$. From a field-theoretical viewpoint, the quantum number $l$ labels the irreducible representations of the group of the rotations induced by the 
coupling of the NACS particle matter field with the non-Abelian gauge field: as a consequence, the values of $l$ are of course quantized and vary over all the non-negative integer and half-integer numbers; $l=0$ corresponds to a system of ideal bosons. As usual, a basis of isospin eigenstates can be labeled by $l$ and the magnetic quantum number $m=-l,-l+1,\cdots,l-1,l$.

The thermodynamics depends in general on the value of the isospin quantum number $l$, the Chern-Simons coupling $\kappa$, and the temperature $T$. In order to enforce the gauge covariance of the theory, the parameter $\kappa$ in (\ref{hamiltonianadelmodello}) has to fulfill the condition $4 \pi \kappa \, = \, {\rm integer}$ \cite{Deser82}. 
Therefore we adopt the notation:
\begin{equation}
4 \pi \kappa \, \equiv k\,.
\label{integer_kappa}
\end{equation} 
Similarly to the Abelian anyons case, the $s-$wave general solution of the radial Schr\"odinger equation (\ref{rsedfmh}), 
derived from the projection of (\ref{hamiltonianadelmodello}) over a generic two-particle isospin channel ($j,j_z$), 
belongs to a one-parameter family accounting for the range of possible boundary conditions, and reads 
\begin{equation} 
\label{nonabelianradialsoftcore} 
R^{j,j_z}_0(r) = \text{const} \cdot \left[ 
        J_{|\omega_j|}(kr) + \sigma \left(\frac{k}{\kappa_{j,j_z}}\right)^{2|\omega_j|}J_{-|\omega_j|}(kr)\,, 
        \right]\,, 
\end{equation} 
where $\sigma=\pm 1$, and $\kappa_{j,j_z}$ is a momentum scale introduced by the boundary condition.  

We refer to the $(2l+1)^2$ quantities \begin{equation}\label{hard-core-par-NA}
\varepsilon_{j,j_z}\equiv \frac{\beta\kappa^2_{j,j_z}}{M}
\end{equation} 
as the hard-core parameters of the system \cite{Mancarella13}, with the hard-core limit corresponding to $\sigma=+1$,
$\varepsilon_{j,j_z}\rightarrow\infty$ for all $j,j_z$.  

We conclude this Section by observing that, according to the regularization used in \cite{Arovas85,Comtet89}, 
the second virial coefficient is defined as  
\begin{equation} 
B_2(\kappa, l, T) - B_2^{(n.i.)}(l,T) = -\frac{2\lambda_T^2}{(2l+1)^2} 
\left[Z^\prime_2 (\kappa, l, T) - Z^{\prime(n.i.)}_2 (l,T)\right]\,\,\,,\label{legamepartizioneviriale} 
\end{equation} 
where $B_2^{(n.i.)}(l,T)$ is the second virial coefficient for the system with particle isospin $l$ and without statistical interaction 
($\kappa \rightarrow \infty$). In Appendix \ref{appendicesoftcore}, $B_2^{(n.i.)}(l,T)$ is expressed in terms of the virial coefficients 
$B_2^B(T)$, $B_2^F(T)$ of the free Bose and Fermi systems with the considered general wavefunction boundary conditions, and 
$\left[Z^{\prime}_2 (\kappa, l, T) - Z^{\prime(n.i.)}_2 (l,T)\right]$ is the (convergent) variation of the divergent partition function 
for the two-body relative Hamiltonian, between the interacting case in exam and the non-interacting limit ($\kappa \rightarrow \infty$).

\section{Internal energy shift for the Lieb-Liniger Bose gas}\label{otherthermod} 

Before studying the energy shift of the Lieb-Liniger model, we consider by comparison the $1d$ hard-core bosons model described by the Hamiltonian:
\be
H_{HC}=-\frac{\hbar^2}{2m}\sum_{i=1}^N\, \frac{\partial^2}{\partial x_i^2}+
\sum_{i<j}\, V_{HC}(x_i-x_j)\,,
\label{HC}
\ee
where 
\be \label{muro}
V_{HC}(x)=\left\{\begin{array}{lcl}
\infty \,\,& , & \text{for }\mid x \mid < a\\
0 \,\, & , & \text{for }\mid x \mid > a\,.
\end{array}\right.
\ee
The thermodynamics of the $1d$ hard-core Bose gas has been determined and studied by Thermodynamics Bethe Ansatz  \cite{Sutherland71,Wadati01}: the relation between pressure and internal energy is a Bernoulli equation 
\cite{Isihara71}
\be
P=\frac{2E}{L(1-a\rho)}\,,
\label{Bern}
\ee
from which it follows that 
\be
E_{res}\,=\, - \frac{PL}{2} \, a\rho\,\,\,.
\label{residual_HC}
\ee
In this case the internal energy shift is negative, due to the fact that the pressure increases for the effect of the  
excluded volume. Furthermore $E_{res}$ vanishes  for $a  \rightarrow 0$, as it should. With regard to the low-dimensional models considered in Sections \ref{themodel}, \ref{otherthermod}, \ref{otherthermod_2d}, the reader will notice that in $2d$ the hard-core condition results in a vanishing internal energy shift, while it does not do likewise in $1d$ (\ref{HC}-\ref{residual_HC}); however, non-hard-core boundary conditions either in $1d$ (\ref{HAM_LL}) and $2d$ (\ref{sol},\ref{nonabelianradialsoftcore}) result in a positive energy shift. Furthermore, unlike the non-hard-core case, the dependence (\ref{residual_HC}) of the internal energy
shift on the temperature is given only by $T$-dependence of the pressure.

Let us now address the internal energy shift of the LL model: using the results of Section \ref{sezioneLiebLiniger}, the pressure and the energy are given by
\be 
\left\{\begin{array}{l}
P = \frac{k_B T}{2 \pi} \int_{-\infty}^{\infty}  \log{\left( 1+e^{-\varepsilon(k')/k_B T} \right)}\,dk'\\
E = L \int_{-\infty}^{\infty} \frac{\hbar^2 k^2}{2m} f(k) dk\,.
\end{array}\right.
\end{equation}

At $T=0$ the energy per particle is given by
\begin{equation}
\frac{E(T=0)}{N}=\frac{\hbar^2}{2m}\,\rho^2\, {\cal E}(\gamma)\,,
\end{equation}
where  ${\cal E}(\gamma)$ is given by 
$${\cal E}(\gamma)=(\gamma/\ell)^3 \int_{-1}^{1} t^2 g(t) dt \,\,\,,$$ 
while the function $g(t)$ is solution of the linear integral equation 
$$g(y)=\frac{1}{2\pi}+\frac{\ell}{\pi} \int_{-1}^{1}\frac{g(t)dt}
{\ell^2+(y-t)^2}$$ 
with $\ell\equiv c/K$ determined from the condition 
$\ell=\gamma \int_{-1}^1  g(t)dt$. It is well known that ${\cal E}\rightarrow 0$ 
for $\gamma \rightarrow 0$ and  ${\cal E}\rightarrow \pi^2/3$ 
for $\gamma \rightarrow \infty$; furthermore ${\cal E} \approx \gamma$ 
for $\gamma \ll 1$ and  ${\cal E}\approx (\pi^2/3) \, (1-4\gamma)$ 
for $\gamma \gg 1$ \cite{Samaj13}.

\begin{figure}[t]
\vspace{0.0cm}
\centerline{\scalebox{0.19}{\includegraphics{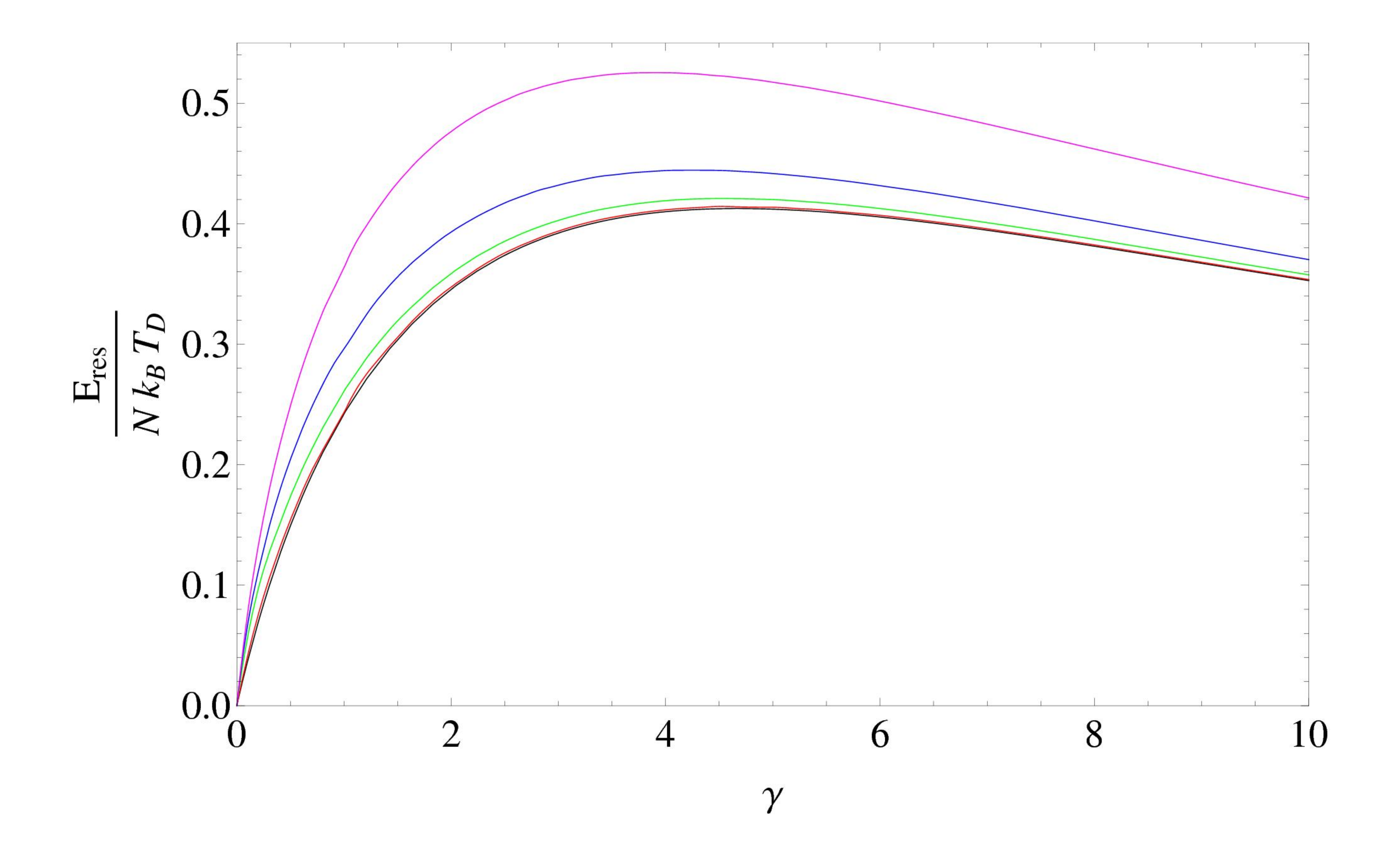}}}
	\caption{Internal energy shift density in units of $k_B  T_D$ as a function 
of the coupling $\gamma$ for different scaled temperatures $\tau$: from bottom 
to top $\tau=0,0.1,0.5,1,2$.}
\label{fig_res_LL}
\end{figure}

At $T=0$ the pressure $P=-(\partial E/ \partial L)_N$ is then
$$
P=\frac{2E}{L} -\frac{\hbar^2\rho^3}{m} \, \frac{\gamma {\cal E}'(\gamma)}{2}\,\,\,,
$$
from which it follows
$$
\frac{E}{N}=\frac{P}{2\rho} + k_B  T_D \, \frac{\gamma {\cal E}'(\gamma)}{2}\,\,\,,
$$
and therefore
\be
\frac{E_{res}}{N k_B T_D}=\frac{\gamma {\cal E}'(\gamma)}{2}\,.
\label{LL_T0_res}
\ee
It is immediately seen that the shift is positive and that it vanishes for $\gamma=0$ ($1d$ ideal Bose gas) and for $\gamma \rightarrow \infty$ (TG gas, having the equation of state of the $1d$ ideal Fermi gas). Furthermore  
\be 
\frac{e_{res}}{k_BT_D} \approx \left\{\begin{array}{lcl}
\frac{\gamma}{2} \,\,& , & \text{for } \gamma \ll 1\\
\frac{2 \pi^2}{3 \gamma} \,\, & , & \text{for }\gamma \gg 1
\end{array}\right.
\ee
and a maximum of the shift appears at a finite value of $\gamma$ (at $\gamma \approx 4.7$). The plot of $e_{res}(T=0)$ in units of $k_B T_D$ is the black lowest curve in Fig.\ref{fig_res_LL}.

At finite temperature one gets
\be
\frac{E_{res}}{Nk_B T_D} = \int_{-\infty}^{\infty} \mathcal{K}^2 f{(\mathcal{K})} 
d\mathcal{K}-\frac{\tau}{4\pi} \int_{-\infty}^{\infty} 
\log{\left(1+e^{-E(\mathcal{K})/\tau}\right)} d\mathcal{K}\,.
\label{LL_res_T}
\ee
A plot of Eq. (\ref{LL_res_T}) as a function of the coupling $\gamma$ for different scaled temperatures is again in Fig.\ref{fig_res_LL}. Even though the shift depends (rather weakly) on the temperature, the same structure occurring at zero temperature is seen: a maximum appears at a finite value of $\gamma$, i.e. between the two ideal bosonic and fermionic limits.

\begin{figure}[t]
\vspace{0.0cm}
\centerline{\scalebox{0.18}{\includegraphics{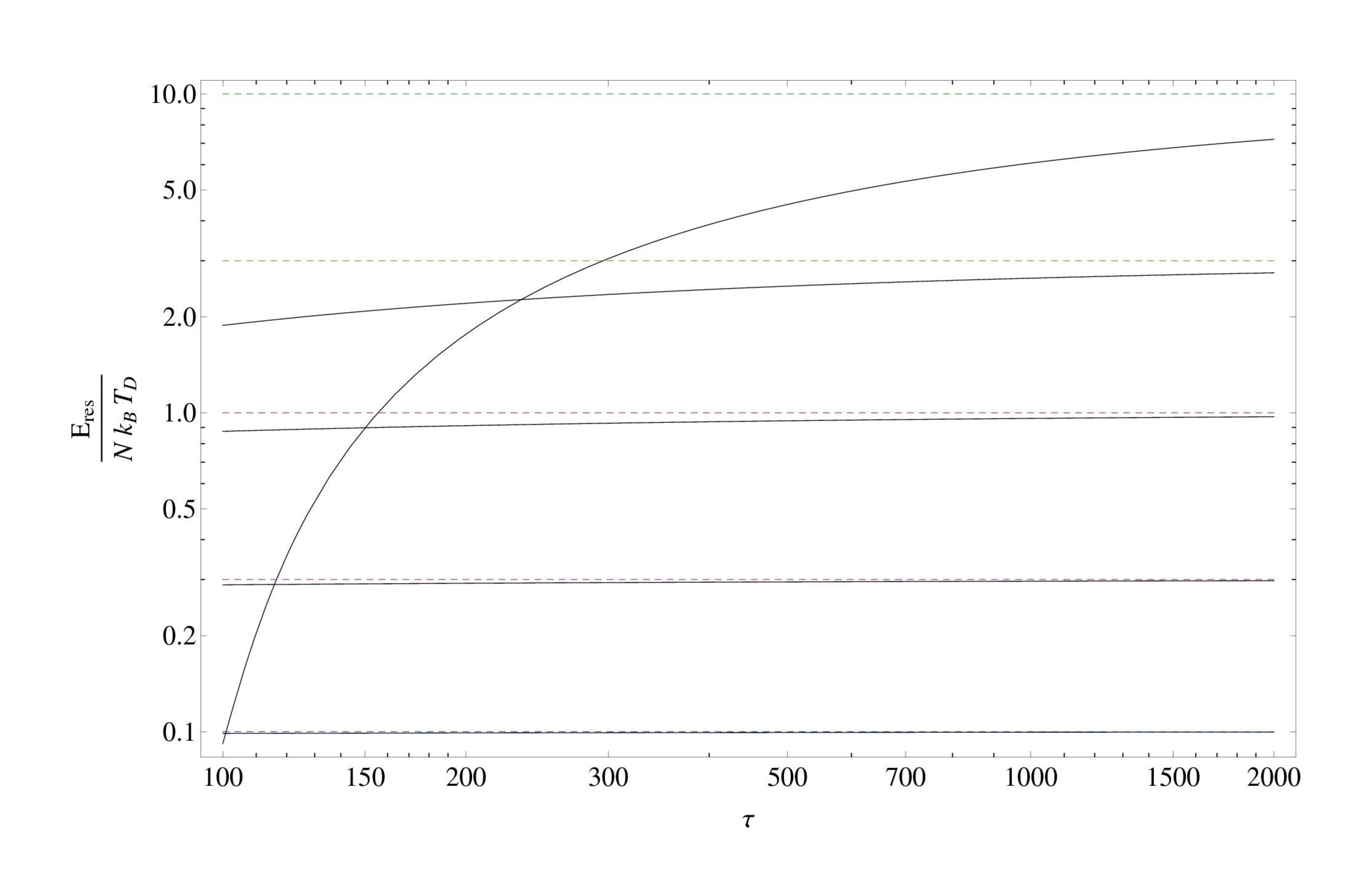}}}
	\caption{Internal energy shift density 
in units of $k_B  T_D$ as a function of the scaled temperature 
$\tau$ for different values $\gamma$ from Eq. (\ref{exp_res_B}): from top 
to bottom it is 
$\gamma=10,3,1,0.3,0.1$ (the dashed line is the asymptotic value $\gamma$).}
\label{fig_res_LL_asympt}
\end{figure}

The high-temperature limit can be explicitly studied: indeed the second virial coefficient \cite{Yang70,Samaj13} written in scaled units is
\be
B_2 = \left\{ \frac{1}{2\sqrt{2}} + e^{\gamma^2/2\tau} \left[ 
\sqrt{\frac{2}{\pi}} \int_{0}^{\sqrt{\gamma^2/2\tau}} e^{-y^2}\,dy - 
\frac{1}{\sqrt{2}}\right] \right\} \lambda_T\,,
\label{B_2LL}
\ee
where $\lambda_T=\sqrt{2\pi \hbar^2/mk_BT}$ is the thermal De Broglie wavelength (\ref{de_B}). Using the virial expansion (\ref{zeroset}) (valid for $\tau \gg 4\pi$) one gets at the first non-trivial order (i.e., $B_2$): 
\be
\frac{E_{res}}{N k_B T} \approx \left( -T \frac{\partial B_2}{\partial T}-\frac{B_2}{2}\right) \, \rho\,.
\label{exp_res}
\ee
Using the relation $\rho \lambda_T=2\sqrt{\pi/\tau}$, from (\ref{B_2LL})-(\ref{exp_res}) one gets then
\be
\frac{E_{res}}{N k_B T_D} \approx \gamma + \sqrt{\frac{\pi}{2\tau}} \, \gamma^2 \,  
e^{\gamma^2/2\tau} \left[ \text{Erf}\left(\sqrt{\frac{\gamma^2}{2\tau}}\right) -1\right] \,,
\label{exp_res_B}
\ee
where we have introduced the error function $\text{Erf}(x)=\frac{2}{\sqrt{\pi}} \int_0^{x} dy \, e^{-y^2}$ \cite{NIST}. 
Using the asymptotic expansion $$\sqrt{\pi}xe^{x^2}[1-\text{Erf}(x)] \approx 1-\frac{1}{2x^2}$$ 
valid for large $x$, one gets
\be 
\frac{E_{res}}{N k_B T_D}=\left\{\begin{array}{l}
\gamma, \,\, \text{for } \gamma^2 \ll 2\tau\\
\frac{\tau}{\gamma}, \,\, \text{for }\gamma^2 \gg 2\tau\,.
\end{array}\right.
\label{esempio1}
\ee
One explicitly sees that $e_{res} \rightarrow 0$ in the two ideal limits $\gamma \rightarrow 0$ and $\gamma \rightarrow \infty$ and that there is maximum between them (roughly $\gamma_{max} \sim \sqrt{\tau}$). From (\ref{esempio1}) we also see that if one fixes a finite coupling $\gamma$ and increases the temperature (i.e., $\tau$), then the internal energy shift density approaches 
the value $k_BT_D\gamma$. Remarkably, the internal energy shift is finite also for infinite temperature: this is shown in Fig.\ref{fig_res_LL_asympt} where $e_{res}/k_B  T_D$ is plotted as a function of the scaled temperature $\tau$ for different values $\gamma$, showing that the asymptotic value $\gamma$ is reached for large temperatures.

\section{Energy-pressure relation for anyonic models}\label{otherthermod_2d}
In this Section we study the Abelian and non-Abelian anyonic gases introduced in Section \ref{sezionescaleinvariant} and we discuss 
their internal energy shift: we show that in the hard-core case the energy-pressure obey (\ref{relazionefondamentale}), therefore in this case the gases have vanishing internal energy shift. The soft-core condition introduces instead a scale and this gives raise to a positive internal energy shift.

In the first part of this Section we treat together the Abelian and non-Abelian gases: the Hamiltonians for Abelian/non-Abelian anyons are defined respectively in (\ref{HNanioniabeliani}) and (\ref{hamiltonianadelmodello}): they are homogeneous with respect 
to the particles coordinates and they scale as 
\be 
H_N(\lambda\textbf{r}_i,\lambda^{-1}\textbf{p}_i)=\frac{1}{\lambda^2}\,H_N(\textbf{r}_i,\textbf{p}_i)\,.
\ee
The hard-core condition at coincident points (in both Abelian and the non-Abelian models) is a particular case 
of a scale--invariant boundary condition, because all finite-$\lambda$-scalings of the $N$-body eigenfunctions
\be \tilde{\psi}(\textbf{r}_i)\equiv \psi(\lambda\textbf{r}_i), \quad \lambda\neq 0
\ee
are hard-core eigenfunctions too, and vanishing whenever any coordinate sits on the boundary of the rescaled volume. Denoting by $A$ the area of the system, in the hard-core case the coordinate scaling results in a dilatation of the energy spectrum:
\be
Sp^{h.c.}[H_N(\lambda^2 A)]=\lambda^{-2}\times Sp^{h.c.}[H_N(A)]\,,
\ee
then the map $(A,\beta)\rightarrow(\lambda^2 A,\lambda^2 \beta)$ 
let $(\log Z)$ invariant. As a consequence of (\ref{relaztermodinamicheperinvarianza}) and 
(\ref{variazionegeneralepartizione}) we obtain the exact identity:
\begin{equation}
E = P\,A\,,
\label{EOS}
\end{equation}  
in agreement with the more general relation (\ref{relazionefondamentale}). 
Equivalently, hard-core anyonic gases fulfill of course:
\be \label{EOS2}
H = 2E\,.
\ee
The validity of (\ref{EOS}) for the particular cases represented by 2D Bose and 2D (spin-less) Fermi ideal gases is remarked in \cite{Khare05}. Thermodynamic relations (\ref{EOS})-(\ref{EOS2}) are not fulfilled by general soft-core NACS ideal gases.
 
We show now that the fulfillment of Eq. (\ref{EOS}) is related to suitable conditions on the virial coefficients. To show it explicitly, 
let us consider now the harmonic regularization of the scale-invariant Hamiltonian for $N$ anyons 
\be H_{N,\omega}=H_N+\omega^2 H_1\,,
\ee
where $H_1\equiv (M/2)\sum\limits_{i=1}^{N} \textbf{r}_i^2$, and $H_N,H_1$ transform according to 
\be 
H_N\rightarrow \frac{1}{\lambda^2}H_N,\quad H_1\rightarrow \lambda^2 H_1 
\label{hamiltscaling}
\ee
under the canonical scaling transformation $(\textbf{r}_i,\,\textbf{p}_i)\rightarrow (\lambda \,\textbf{r}_i,\,
\frac{1}{\lambda} \,\textbf{p}_i)$; as a consequence, the following relation holds for the regularized Hamiltonian, 
expressed in terms of scaling for the regularizing frequency and the spatial coordinates:	 
\be 
  H_{N,\gamma\,\omega}(\frac{1}{\sqrt{\gamma}}\,\textbf{r}_i,\,\sqrt{\gamma}\,\textbf{p}_i)=\gamma \, H_{N,\omega}(\textbf{r}_i,\,\textbf{p}_i)\;,\quad \forall \gamma \neq 0\,.
\ee
Notice that the harmonic regularization breaks the scale--invariance, which is retrieved in the $\omega \rightarrow 0$ limit.

Now we apply the hard-core condition. For any eigenfunction $\psi_n(\textbf{r}_i)$ of $H_{N,\omega}(\textbf{r}_i,\,\textbf{p}_i)$ 
fulfilling the hard-core boundary condition, we correspondingly get 
$\tilde{\psi}_n(\textbf{r}_i)\equiv \psi_n(\sqrt{\gamma}\,\textbf{r}_i)$ 
(also fulfilling the hard-core boundary condition) 
as eigenfunction of $H_{N,\gamma\,\omega}(\textbf{r}_i,\,\textbf{p}_i)$, 
so that, denoting the hard-core condition by the superscript ''\textit{h.c.}'', the frequency acts barely as a dilatation for the energy spectrum:
\be
Sp[H_{N,\gamma\,\omega}^{h.c.}]=\gamma\times Sp[H_{N,\omega}^{h.c.}]\,,
\label{formulaspettro}
\ee
whence the $N$-body partition function
\be 
Z_{N}^{h.c.}(\beta,\omega)=\Tr e^{-\beta H_{N,\omega}^{h.c.}}
\ee
fulfills $\forall x \in \mathbb{R}$
\be 
Z_{N}^{h.c.}(\beta,\omega)=Z_{N}^{h.c.}(\beta/x,x\omega)\,.
\label{scalingpartition}
\ee

Assuming  valid the existence of the virial expansion, one has that the thermodynamic relations 
\be 
\left\{\begin{array}{l}
\mathcal{L}=\sum\limits_{N=0}^\infty z^N\,Z_N\,,\\
P=\frac{k_BT}{A} \ln \mathcal{L}\,,\\
P= \rho k_B T \,\sum\limits_{N=0}^\infty B_{N+1}(T)\,\rho^N \,,\\
\rho=z\,\frac{\partial}{\partial z}\left(\frac{1}{A} \ln \mathcal{L}\right)\vert_{A,T}\,
\end{array}\right.
\ee
imply that the coefficients $B_{N+1}(T)\lambda_T^{-2N}$ of the pressure expansion in powers of the dilution parameter $\rho \lambda_T^2$ can be expressed as a rational combination $f_N(\{Z_i(\beta,\omega)\})$ 
of the first partition functions up to $Z_{N+1}$. The assumed existence of virial expansion, 
together with scaling (\ref{scalingpartition}), enforces
\be 
B_{N+1}(T)\lambda_T^{-2N}=\lim_{\omega\rightarrow 0}f_N(\{Z_i(\beta,\omega)\})=\lim_{\beta'\rightarrow \infty}f_N(\{Z_i(\beta',\omega=0)\})\,,
\ee 
so that $B_{N+1}(T)\lambda_T^{-2N}$ has to be independent of temperature, thus hard-core ideal anyonic gases fulfill 
\be 
B_{k+1}(T)\propto T^{-k}\,,\label{Tindephardcorevir}.
\ee 
From this relation follows that for these systems the last three identities of (\ref{zeroset}) take the form  
\be \left.\begin{array}{l}
\text{Entropy}: \,\,\,\,\,\, \frac{S}{Nk_B}=2-\log(\rho\lambda_T^2)+\sum\limits_{k \geq 1} \frac{k-1}{k}\,  B_{k+1}\, \rho^k\,;\\ 
\text{Internal energy}: \,\,\,\,\,\, \frac{E}{Nk_BT}=
1+\sum\limits_{k \geq 1} B_{k+1}\,\rho^k\,;\\
\text{Enthalpy}: \,\,\,\,\,\, \frac{H}{Nk_BT}=2+2\sum\limits_{k \geq 1} B_{k+1}\,\rho^k\,.\end{array} \right\}\text{(hard-core case)}
\label{secondoset}
\ee

We point out that the corresponding entropy and heat capacity at constant volume are unaffected by the statistical interaction at 
the lowest order of virial expansion (being independent of $B_2$). Using formula (\ref{generalformulasoftcorecase})
for $B_2$, one can obtain the leading deviation of the various thermodynamic quantities from their ideal gas value.

An important consequence of (\ref{Tindephardcorevir}) is that from (\ref{zeroset}) and (\ref{secondoset}) one gets again 
Eq.(\ref{EOS})
at all orders of the virial expansion for hard-core Abelian and non-Abelian anyonic gases (within  the convergence radii  of these expansion) \cite{note}, in agreement with the general relation \ref{relazionefondamentale}

The scaling properties for isoentropic processes derived in Sec. \ref{sezionescaleinvariant} apply in particular to 
Abelian and non-Abelian anyonic gases with hard-core conditions. Isoentropic processes between initial and final states 
at equilibrium of hard-core anyonic gases are characterized by the following relation between internal energy and temperature:
\be 
E\propto A^{-1}\,;\quad T\propto A^{-1}
\label{trasformazioneisoentropicaanioni}
\ee
As a consequence, for hard-core anyonic gases subjected to an isotropic transformation one gets $P\propto A^{-2}$, in agreements with (\ref{EOS}) and (\ref{trasformazioneisoentropicaanioni}), and the dilution parameter $x \equiv \rho \lambda_T^2$ remains invariant along isoentropic curves. Furthermore, according to (\ref{proporzenergiatemperatura}), the internal energy associated with equilibrium states of an isoentropic process takes the form $E=y\times Nk_BT$, where $y$ remains constant along the isoentropic curve, 
while it depends solely on the dilution parameter $x=\rho\,\lambda_T^2$ over the entire phase diagram. Since we are in two dimensions, $y$ is just the compressibility factor. 

Remarkably, Eq. (\ref{proporzenergiatemperatura}) traces the study of free anyonic thermodynamics back to the determination of how the compressibility factor $y(x)$ depends on the dilution parameter $x$, and this is a genuine consequence of the scaling symmetry, 
valid therefore also beyond the radius of convergence of the virial expansion. 

For the family of systems represented by Abelian anyons gases, where $\alpha$ will denote henceforth the statistical parameter 
as in Subsection \ref{abeliananyons}, the factor $y$ can be parametrized as $y=y(x,\,\alpha)$. The cases $y(x,0)$ (2D Bose gas) and $y(x,1)$ (2D Fermi gas) can be traced back to the analysis in Chapter 4 of Ref.\,\cite{Khare05}; as the dilution parameter $x$ is swept from $0$ to $\infty$ the gas moves from ideality to an increasingly dense regime, and $y(x,0)$ monotonically decreases from $1$ to $0$, while $y(x,1)$ monotonically increases from $1$ to $\infty$; low/high density limit behaviors immediately follow from \cite{Khare05}:
\be 
\left\{\begin{array}{lr}y(x,0)\sim 1-\frac{x}{4}+\sum\limits_{l=1}^\infty\,\frac{x^{2l}}{(2l+1)!}B_{2l},\quad & x \ll 1 \\
y(x,0)\sim \frac{\pi^2}{6}x^{-1},\quad & x \gg 1\end{array}\right.
\,, \quad\quad
\left\{\begin{array}{lr}y(x,1)\sim 1+\frac{x}{4}+\sum\limits_{l=1}^\infty\,\frac{x^{2l}}{(2l+1)!}B_{2l},\quad & x \ll 1 \\
y(x,1)\sim \frac{x}{2},\quad & x \gg 1\end{array}\right.\,,
\ee
where $B_{n}$ denote here the Bernoulli numbers \cite{NIST}. As expected, the general behavior at intermediate $\alpha$ is non-trivial, while the basic qualitative statements about $y(x,\alpha)$ are that $y(0,\alpha)=1$ for any $\alpha$ (limit of ideal gas) and
\be 
y(x,\alpha)\sim 1-\frac{1}{4}(1-4\alpha+2\alpha^2)\,x,\quad x\ll 1\,.
\ee
since the dominance of the second virial coefficient in a very dilute regime. This approximate behavior interpolates the curves $y(x,0)$ and $y(x,1)$, and the sign of its slope at $x=0$ switches at $\alpha=1-\sqrt{1/2}$, i.e., within the dilute regime approximation the statistical energy is negative for $0 \leq \alpha < 1-\sqrt{1/2}$, positive for $1-\sqrt{1/2}<\alpha \leq 1$.  

A remarkable perturbative result is argued in Eq.(22) of \cite{Comtet91} about the ground state energy for Abelian anyons, which, 
by assuming the continuity of $E(N,A,T)$ at $T=0$, reads here 
\be
y(x,\alpha)\sim \frac{\alpha}{2}\,x, \quad x\gg 1 \;\text{ and }\; \alpha \ll 1\,.
\ee

Let us pause here for a comment about the classical limit of the hard-core anyonic system. In the picture of anyons as charge-flux composites written for instance in the bosonic bases, one is free to consider arbitrarily large magnetic fluxes $\Phi=\alpha\,h/q$, 
$q=$ being the charge of the particles. The kinetic terms alone would yield the Bose statistics for the quantum case, and the Boltzmann statistics for its classical limit. The quadratic terms in $\alpha$ should be regarded as self-energies of the vortices (in both cases). Finally, the momentum-flux terms correspond to the magnetic inter-particle interaction, and they are responsible for the non-trivial anyonic thermodynamics, which is periodic in the flux variable $\alpha=q\Phi/h$. Correspondingly, in the classical limit 
the magnetic vector potential does not affect the motion of particles at all, or, in other words, the Aharonov-Bohm effect disappears for classical charges, and the gas would approach the classical ideal gas law, no matter how large the fluxes attached to the particles are. 
Anyonic thermodynamics is intrinsically a quantum one, ruled solely by the dilution parameter $x=\rho \lambda_T^2$ 
because no additional length scale (besides the thermal length) is set by the flux parameter. The vanishing of the dilution parameter in the classical limit leads again (from a different point of view) to a trivial thermodynamics regardless of the size of fluxes. 
Finally, from the viewpoint of density of states, we may think about the effect of the statistical interaction in terms of what happens, e.g., to the $2$-anyon spectrum in an harmonic trap $(1/2)m\, \omega_0^2\, r^2$. This interaction acts \cite{Khare05} 
as a uniform upward spectral shift whose $\alpha$-dependence is periodic with finite period $\Delta \alpha=2$, 
therefore this shift becomes irrelevant (to the density of states) for any values of the flux in the classical limit $k_B T\gg \hbar \omega_0$.

\subsection{Soft-core anyons}  
The scale--invariance in force for hard-core anyons does not apply in presence of soft-core boundary conditions, in which case we will compare internal energy and pressure within the dilute regime (up to the first order in the dilution parameter $\rho\lambda_T^2$). Let us define the relative internal energy shift density $e_{rel}$ 
as the dimensionless quantity
\be 
e_{rel}\equiv\frac{E-PA}{Nk_BT}= -\rho\left( B_2^{s.c.}+T\frac{d}{dT}B_2^{s.c.}\right)+O((\rho\lambda_T^2)^2)\,.
\label{internalenergyshift}
\ee

\begin{figure}[t]
\vspace{0.0cm}
\centerline{\scalebox{0.15}{\includegraphics{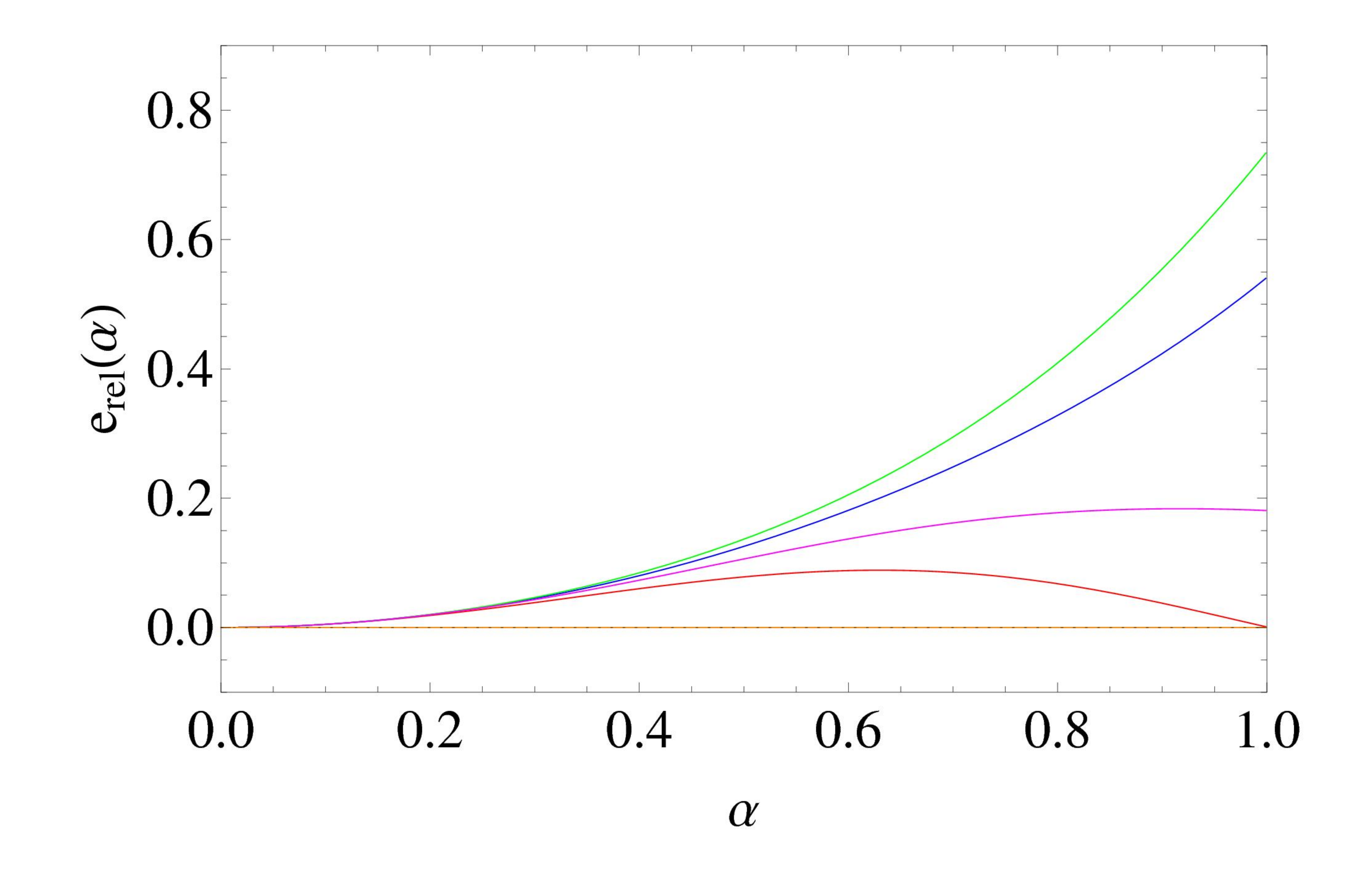}}}
	\caption{Leading term of the relative energy shift 
$e_{rel}(\alpha,\varepsilon,T)$ in units of $\rho \lambda_T^2$ 
as a function of the statistical parameter $\alpha$, for different values of the hard-core parameter for Abelian anyons. 
The $5$ curves are obtained for $\sigma=1$: from top to bottom, $\varepsilon=$ $1$ (green), $2$ (blue), 
$0.1$ (magenta), $10$ (red), $\infty$=hard-core or $0$ (orange). The relative energy shift is non-negative, 
vanishing at the bosonic point $\alpha=0$, and reaches its minimum at $\varepsilon=\infty, 0$ 
(where $e_{rel}$ vanishes for any $\alpha$). $ e_{rel}(\alpha,\varepsilon,T)$ is periodic in 
$\alpha$ with period $2$, and is symmetric with respect to all the integer value of  $\alpha$; 
for soft-core cases ($\varepsilon>0$), $e_{rel}$ has in general cusps in $\alpha$ at the fermionic points.}
\label{energyshift_vs_stat_param_softcore}
\end{figure}

For Abelian anyons, Eqs. (\ref{explicitintegralform})-(\ref{internalenergyshift}) give
\be 
e_{rel}=2\rho \lambda_T^2\,T\frac{d}{dT}f(T) +O((\rho\lambda_T^2)^2)\,,
\ee
where
\be 
 f(T)\equiv e^{\varepsilon(T)} \theta(-\sigma)  
+\frac{\alpha\sigma}{\pi} \sin{\pi\alpha} 
\int_0^\infty \frac{dt\, e^{-\varepsilon(T) t}\, t^{|\alpha|-1}}{1+2\sigma\cos{\pi\alpha}\;t^{|\alpha|}+t^{2|\alpha|}}  \,,\quad \varepsilon(T)=\frac{\kappa^2}{Mk_BT}\,.
\ee
The resulting shift is
\be 
e_{rel}(\alpha,T,\varepsilon)=2\rho \lambda_T^2\,\varepsilon\,\left[ -e^{\varepsilon} \theta(-\sigma)  
+\frac{\alpha\sigma}{\pi} \sin{\pi\alpha} 
\int_0^\infty \frac{dt \,e^{-\varepsilon t} \, t^{|\alpha|}}{1+2\sigma\cos{\pi\alpha}\;t^{|\alpha|}+t^{2|\alpha|}} \right]+O((\rho\lambda_T^2)^2)\,,
\label{abelianshift}
\ee
whose leading term in $\rho \lambda_T^2$ is illustrated in Fig.\,\ref{energyshift_vs_stat_param_softcore}. The plot of the shift $e_{rel}(\alpha,T,\varepsilon,\sigma=1)$ exhibits a smooth behavior in the bosonic points and a cusp in the fermionic ones, as soon as the hard-core condition is relaxed. We observe that the restriction of $e_{rel}(\alpha,T,\varepsilon)$
over the interval $\alpha \in [0,1]$ is not a monotonic function of $\alpha$ for any $\varepsilon$. The proportionality $E=PA$ remains valid at the bosonic points also in the soft-core case [i.e. $e_{rel}(\alpha=2n,T,\varepsilon)=0$], and the monotonicity of the shift $e_{rel}$ as a function of $\alpha$ occurs for any $\varepsilon \in [\varepsilon_{-},\varepsilon_{+}]$,\; $\varepsilon_{-}\approx 0.13$, $\varepsilon_{+}\approx 3.0$, and in particular the relative shift is maximal at the fermionic points for any $\varepsilon \in [\varepsilon_{-},\varepsilon_{+}]$, 
while outside of this interval the shift due to the soft-core boundary conditions reaches its maximum at an internal point $\alpha_{max}(\varepsilon)$, featuring the properties $\alpha_{max}(\varepsilon)\rightarrow 1^-$ 
for $\varepsilon\rightarrow (\varepsilon_{\pm})^{\pm}$, and 
$\alpha_{max}(\varepsilon)\rightarrow 0^+$ for 
$\varepsilon\rightarrow 0$ or $\varepsilon\rightarrow\infty $ .

\begin{figure}
       \includegraphics[width=0.45\linewidth]{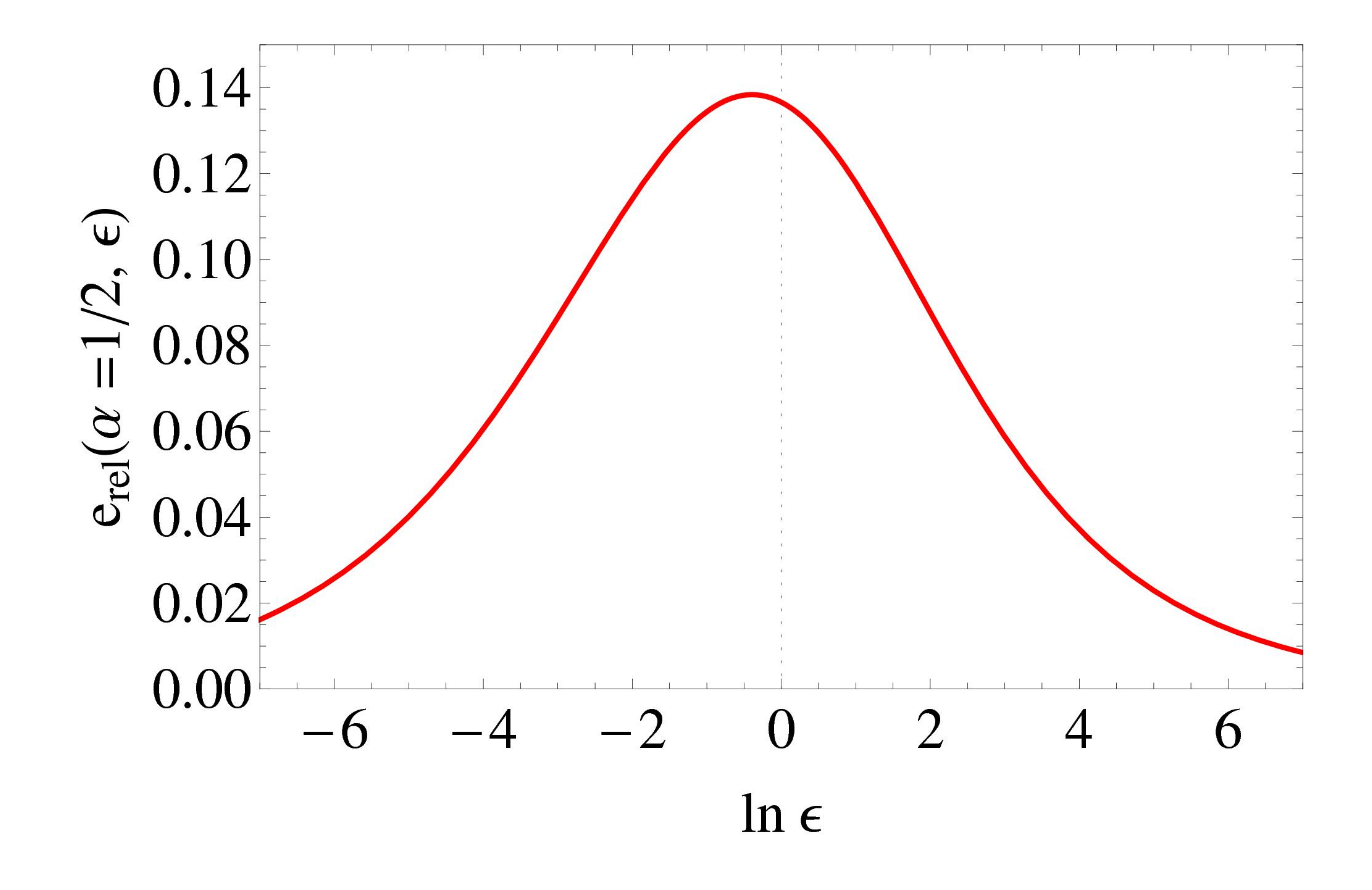}
       \includegraphics[width=0.45\linewidth]{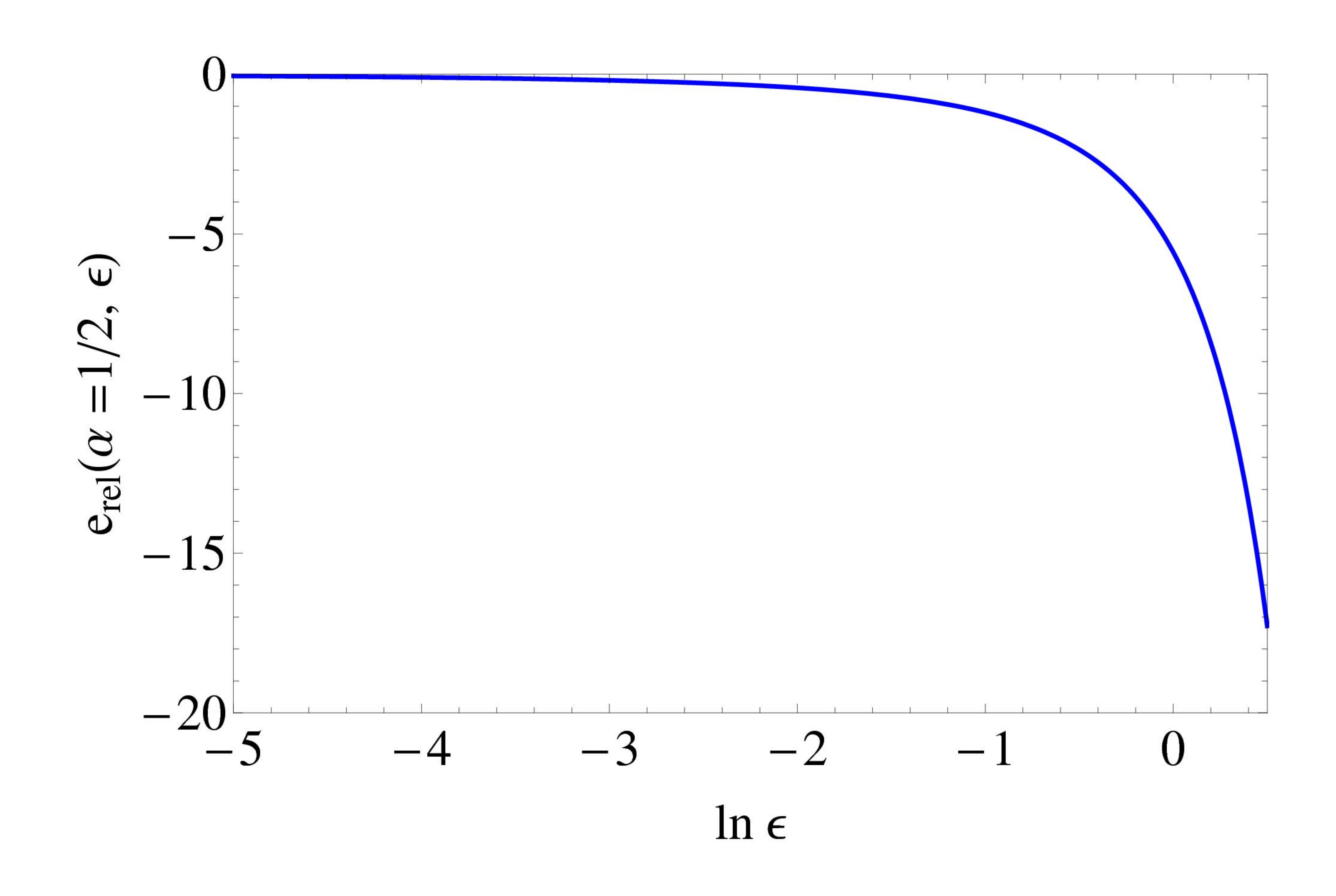}
        \caption{Relative energy shift 
$e_{rel}(\alpha=1/2,\sigma,\varepsilon,T)$ for semions 
($\alpha=integer\,+1/2$) 
in units of $\rho \lambda_T^2$ 
as a function of the hard-core parameter in logarithmic scale; 
$\sigma=+1,-1$ in the left/right panel. 
Left: the relative shift 
$e_{rel}(\alpha=1/2)$ is positive for $\sigma=1$, 
vanishing in the scale-invariant limits 
$\varepsilon=0,\infty$, and reaches 
its maximum  $\approx 0.14$ about $\varepsilon$ of the order of unity.
Right: there is a bound state in the energy spectrum 
for $\sigma=-1$, and the corresponding 
relative energy shift is negative and 
rapidly tending to $-\infty$ as $\varepsilon$ increases.} 
        \label{figshift_vs_softcoreparameter}
\end{figure}

As a particular case, the following expressions, plotted in Fig.\,\ref{figshift_vs_softcoreparameter} for $\sigma=\pm 1$,  hold 
for the semion case ($\alpha=\text{integer}+1/2$):
\be 
e_{rel}(\alpha=1/2,T,\varepsilon,\sigma=+1)\left\{\begin{array}{l}=\rho \lambda_T^2\,\left(\sqrt{\frac{\varepsilon}{\pi}}-\varepsilon\,e^\varepsilon\,\erfc(\sqrt{\varepsilon})\right)\,+O((\rho\lambda_T^2)^2)\\
\sim \rho \lambda_T^2 \sqrt{\frac{\varepsilon}{\pi}},\quad \varepsilon \rightarrow 0\\
\sim \rho \lambda_T^2 \frac{1}{2\sqrt{\pi \varepsilon}},\quad \varepsilon \rightarrow \infty\end{array}\right.
\,,
\ee
and 
\be 
e_{rel}(\alpha=1/2,T,\varepsilon,\sigma=-1)\left\{\begin{array}{l}=\rho\lambda_T^2\left(\varepsilon\,e^\varepsilon
\left[\erfc(\sqrt{\varepsilon})-2\right]-
\sqrt{\frac{\varepsilon}{\pi}}\right)\,+O((\rho\lambda_T^2)^2)\\
\sim -\rho \lambda_T^2 \sqrt{\frac{\varepsilon}{\pi}},\quad \varepsilon \rightarrow 0\\
\sim -2\, \rho \lambda_T^2 \, \varepsilon \, e^\varepsilon,\quad \varepsilon \rightarrow \infty\end{array}\right. \,.
\ee

\begin{figure}[t]
\vspace{0.0cm}
\centerline{\scalebox{0.15}{\includegraphics{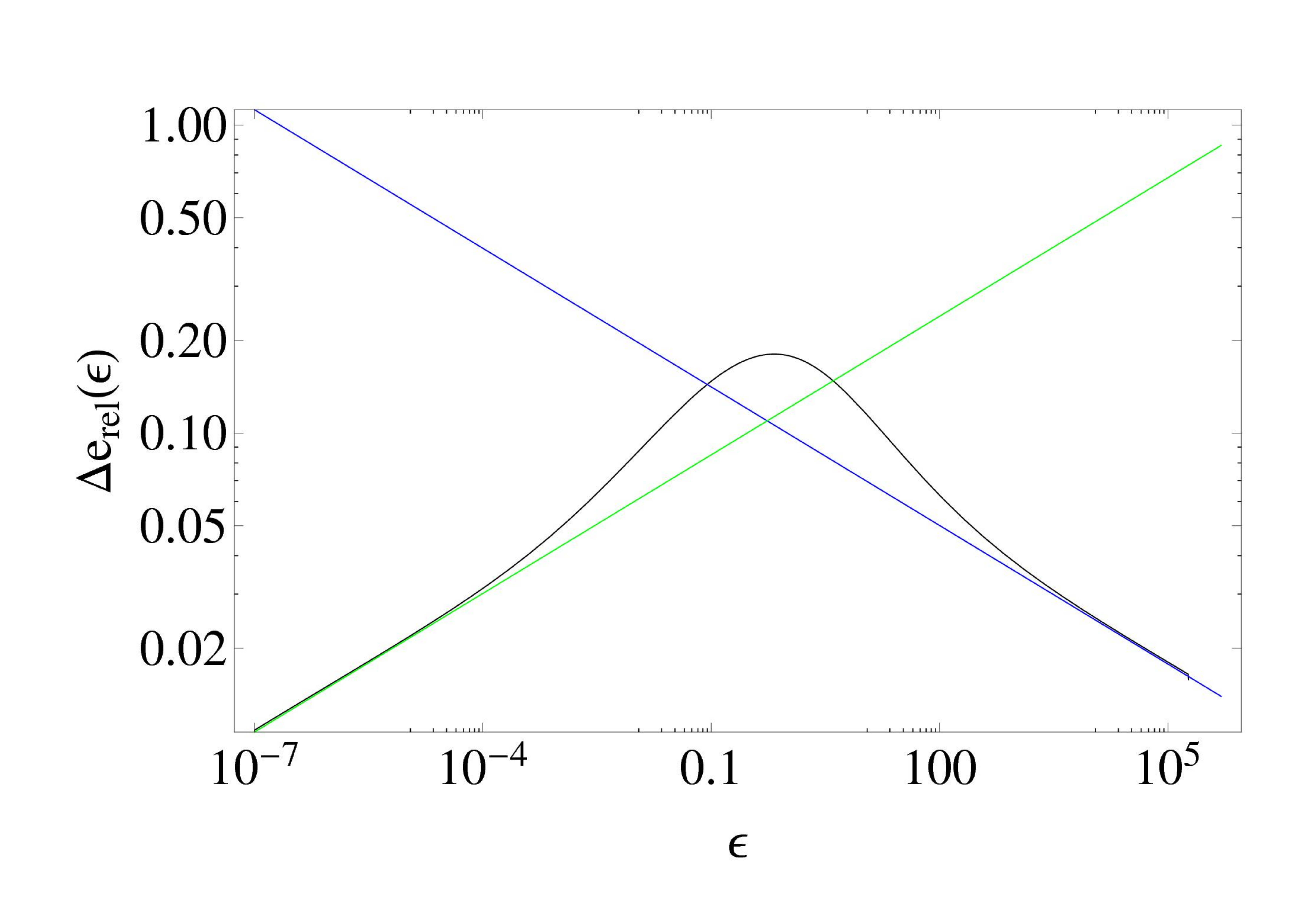}}}
	\caption{In black, $e_{rel}(k=3,l=1/2,\varepsilon,\sigma=+1,T)$ 
in units of $\rho \lambda_T^2$ as a function of the
 hard-core parameter $\varepsilon$ in log-log scale, for $k=3$ and $l=1/2$, in the completely isotropic case for non-Abelian gas 
($\varepsilon_{0,0}=\varepsilon_{1,m}=\varepsilon$, for $m=1,0,-1$). 
In green and blue, small-$\varepsilon$ and large-$\varepsilon$ 
asymptotic behaviors 
(respectively $ \approx 0.12 \,\varepsilon^{0.15}$ and 
$\approx 0.1\, \varepsilon^{-0.15}$).}
\label{energyshiftnonAbel_vs_stat_param_softcore}
\end{figure}

Similar  expressions can be  worked  out for the non-Abelian case: for simplicity, we only consider the completely isotropic case 
$\varepsilon_{jj_z}\equiv \varepsilon$, whose relative energy shift defined above, by virtue of (\ref{simplifiedsoftcore})-
(\ref{modifiedparameter})-(\ref{internalenergyshift})-(\ref{abelianshift}), 
is 
\be 
e_{rel}(\kappa,l,T,\varepsilon,\sigma)=\frac{1}{(2l+1)^2}\sum_{j=0}^{2l}(2j+1)\,e_{rel}(\nu_j,T,\varepsilon,\sigma),\quad \; \nu_j\equiv\left(\omega_j-\frac{1+(-1)^{j+2l}}{2}\right) \, mod \, 2-1\,\,\,.
\label{softcorenergyshift}
\ee
Note that, in a soft-core dilute NACS gas, the internal energy has always a finite relative shift from its ideal-gas value $E=PA$; in particular, the shift (\ref{softcorenergyshift}) has the same sign of the parameter $\sigma=\pm 1$, 
due to the positivity of the denominator ($1+2\sigma\cos{\pi\alpha}\;t^{|\alpha|}+t^{2|\alpha|}$) 
in (\ref{abelianshift}); indeed, the $2l+1$ distinct $j$-channels 
give contributions of equal sign to (\ref{softcorenergyshift}).
The dependence of the relative energy shift on $\varepsilon_{j j_z}\equiv \varepsilon$ is illustrated in 
Fig.\,\ref{energyshiftnonAbel_vs_stat_param_softcore} 
for the case $4\pi \kappa\equiv k=3$, $l=1/2$, $\sigma=+1$. 
The figure clearly shows a power-law decay to zero of $e_{rel}/(\rho\lambda_T^2)$, as the hard-core parameter $\varepsilon$ 
approaches very small/large values (in agreement with the scale--invariance of these two limit cases): 
$e_{rel}(k=3,l=1/2,\varepsilon,T)/(\rho\lambda_T^2) \approx 0.12\, 
\varepsilon^{0.15}$ for $\varepsilon \ll 1$, 
while $e_{rel}(k=3,l=1/2,\varepsilon,T)/(\rho\lambda_T^2) 
\approx 0.1\, \varepsilon^{-0.15}$ for $\varepsilon \gg 1$.
The vanishing of $e_{rel}/(\rho\lambda_T^2)$ in the scale--invariant 
limit cases is asymptotically approached, although 
for very large/small $\varepsilon$: for $\varepsilon=10^{\pm 5}$, $e_{rel}(k=3,l=1/2,\varepsilon,T)/(\rho\lambda_T^2)$ still deviates 
from its asymptotic value zero by $\approx 2\times 10^{-2}$ 
in both cases. Therefore even a tiny deviation from scale--invariance 
may still have an impact on the thermodynamics of this family of non-Abelian anyons. 

\begin{figure}[t]
\vspace{0.0cm}
\centerline{\scalebox{0.15}{\includegraphics{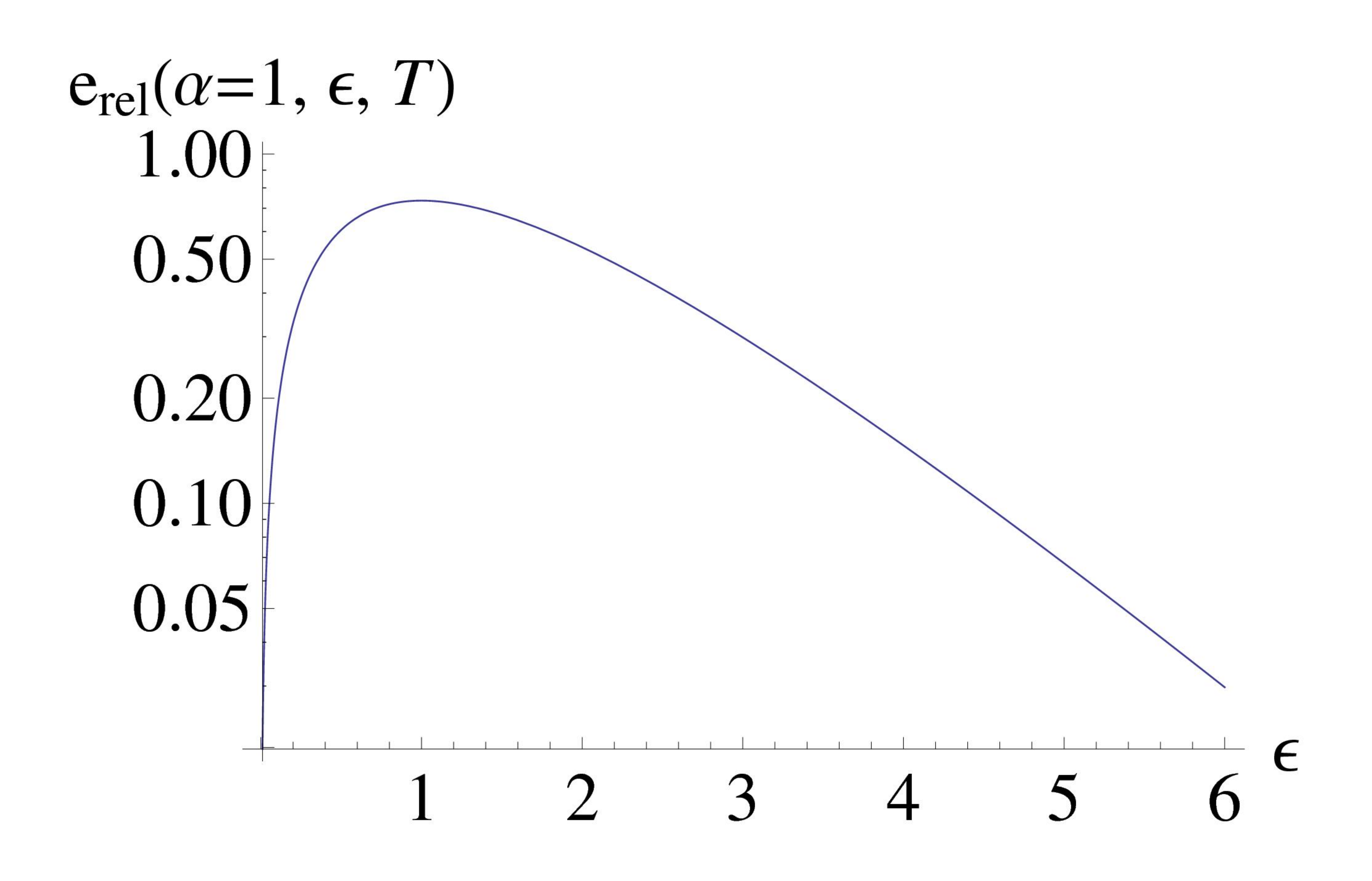}}}
	\caption{$e_{rel}(\alpha=\pm 1,\varepsilon,\sigma=+1,T)$ in units of 
$\rho \lambda_T^2$ as a function of the hard-core parameter $\varepsilon$ 
in logarithmic scale.}
\label{expondecayabeliananyon}
\end{figure}

This feature can be contrasted with the decay of $e_{rel}(\alpha=\pm 1,\varepsilon,T)$, 
as $\log \varepsilon \rightarrow \pm\infty$, for Abelian anyons in the fermionic limit: 
in this case, the energy shift is a power law in $\varepsilon$ (and precisely linear) for $\varepsilon \ll 1$, 
while instead it decays exponentially in $\varepsilon$ for $\varepsilon \gg 1$, as seen in 
Fig.\ref{expondecayabeliananyon}. The shift $e_{rel}(\alpha=\pm 1,\varepsilon,T)$ reaches 
its maximum $2/e\approx 0.74$ (in units of $(\rho\lambda_T^2)$) at $\varepsilon=1$.

\section{Conclusions}
\label{conclusions}
In this paper we have shown that a particularly simple relation of proportionality between internal energy and pressure holds 
for scale--invariant thermodynamic systems (with Hamiltonian homogeneous function of the coordinates), including 
classical and quantum -- Bose and Fermi -- ideal gases. To quantify the deviation from such a relation we have introduced and studied the internal energy shift as the difference between the internal energy of the system and the corresponding value $E=(d/\alpha)PA$ 
of the internal energy for scale--invariant systems. This internal energy shift is a kind of ``departure'' internal energy, where in general a departure function measures deviations from the ideal gas  behavior. The internal energy shift defined in (\ref{relazionefondamentale}) measures a deviation from the ideal gas by keeping $V$ and $P$ fixed: this has to be compared with the conventionally defined departure (or residual) internal energy (defined for fixed $P$ and $T$). The latter is not zero for general scale--invariant systems (and in particular it is not vanishing for hard-core ideal anyons): one can indeed see that if a system is scale--invariant and the conventionally defined departure internal energy is zero, then it has to be an ideal gas. At variance, our definition (\ref{defresidualenergy}) of the internal energy shift is vanishing for all scale--invariant systems, including hard-core anyons for which we show that $E=PA$.  An example of system which is scale--invariant, but having non-vanishing conventionally-defined departure internal energy is the hard-core anyonic gas, as discussed in Section \ref{otherthermod_2d}: summarizing, the quantity (\ref{defresidualenergy}) can be regarded as a good measure of the deviation from scale--invariance for non-ideal gases.

In particular, we have provided criteria for which the internal energy shift density of an imperfect (classical or quantum) 
gas is a bounded function of temperature. We have also shown that for general scale--invariant systems the dependence of virial coefficients upon the temperature is very simple, and is expressed by Eq. (\ref{relation_B_T}).

We have considered deviations from the energy-pressure proportionality in low-dimensional models of gases which interpolate between the ideal Bose and the ideal Fermi gases,  focusing the attention on the Lieb-Liniger model in $1d$ and on the anyonic gas in $2d$. In $1d$ the internal energy shift is determined from the thermodynamic Bethe ansatz integral equations and an explicit relation for it is provided at high temperature: the internal energy shift is positive, it vanishes in the two limits of zero 
and infinite coupling (respectively the ideal Bose and the Tonks-Girardeau gas) and it has a maximum at a finite, temperature-depending, value of the coupling. Remarkably, at fixed coupling the internal energy shift density saturates to a finite value for infinite temperature. 

In $2d$ we have considered systems of Abelian anyons and non-Abelian Chern-Simons particles and we have showed 
that the relation between the internal energy and the pressure of anyonic gas is exactly the same found for $2D$ Bose and Fermi 
ideal gases as long as the hard-core case is considered. Soft-core boundary conditions introduce a length scale and determine a 
non-vanishing internal energy shift: we have provided details about this shift in the dilute limit.
Asymptotic expressions with respect to the hard-core parameter $\varepsilon$ are derived for both Abelian and non-Abelian soft-core anyonic gases.

\vspace{3mm}
{\it Acknowledgements:} We thank Thors Hans Hansson and Marton Kormos for discussions, and Yoseph Stein for valuable
discussions during the early stages of the work. This study has been supported by Nordita Grant number ERC 321031-DM and VR VCB 621-2012-2983.

\appendix
\section{Proof of conditions on the boundedness of internal energy shift}
\label{condizioneenergiaresidua}
We use Eqs. (\ref{zeroset}) in order to write 
the internal energy shift in the dilute limit for quadratic 
dispersion relation of the particles
\be 
e_{res}=\frac{E-\frac{d}{2}PV}{N}=\frac{1}{\beta}\left(\frac{E}{Nk_BT}-\frac{\frac{d}{2}PV}{Nk_BT}\right)=\left(\frac{\partial f(x)}{\partial x}-\frac{d}{2}\frac{f(x)}{x}\right)\rho\,,
\ee
where $f(x)$ denotes the second virial coefficient 
as a function of its unique variable $x \equiv 1/(k_B T)$. 
We are interested in the boundedness of $e_{res}$ 
as the high-temperature limit $x \rightarrow 0$ is approached, i.e.
\be
\lim\limits_{x\rightarrow 0} \left[\partial_x f(x)-\frac{d}{2}\frac{f(x)}{x}\right]= const.
\ee
By setting $g(x)\equiv f(x)/x^{d/2}$, the above condition requires $g(x)=c_1'+c_2'\,x^{1-(d/2)}+o(x^{1-(d/2)})$ for any $d\neq 2$, and $g(x)=c_1'+c_2' \log x+o(\log x)$ for $d=2$, with $c_i'$ arbitrary constants, therefore 
\be
\left\{
\begin{array}{ll} 
f(x)=c_1\,x^{d/2}+c_2\,x+o(x)\;,& \text{ for } d\neq 2\\
f(x)=c_2\,x \log x+o(x \log x)\;,& \text{ for } d=2
\end{array}
\right.,
\ee
from which criteria (\ref{criterio1})-(\ref{criterio2})-(\ref{criterio3}) 
follow.

\section{NACS quantum statistical mechanics}\label{nacsqsm}

The interaction terms in $H_N$ in the NACS model can be removed by a similarity transformation:  
\begin{eqnarray} 
{H}_N&\longrightarrow & UH_N U^{-1}= H^{\rm free}_N = -\sum^N_\alpha 
\frac{2}{M_\alpha} \partial_{\bar z_\alpha}\partial_{z_\alpha}\nonumber\\ 
\Psi_H &\longrightarrow & U \Psi_H = \Psi_A \label{simil} 
\end{eqnarray} 
where $U(z_1,\dots,z_N)$ satisfies the Knizhnik-Zamolodchikov (KZ) 
equation (\cite{kz}): 
\begin{equation} 
\left(\frac{\partial}{\partial z_\alpha}  - \frac{1}{ 2\pi 
\kappa} \sum_{\beta\not=\alpha} \hat Q^a_\alpha \hat Q^a_\beta \frac{1}{z_\alpha -z_\beta}\right) U(z_1,\dots,z_N) =0\,\,\,, 
\end{equation} 
and $\Psi_H(z_1,\dots,z_N)$ stands 
for the wavefunction of the $N$-body system of 
the NACS particles in the holomorphic gauge. 
$\Psi_A(z_1,\dots,z_N)$ obeys the braid statistics \cite{Lee93} 
due to the transformation function $U(z_1,\dots,z_N)$, 
while $\Psi_H(z_1,\dots,z_N)$ satisfies ordinary statistics: 
$\Psi_A(z_1,\dots,z_N)$ is commonly referred to as 
the NACS particle wavefunction in the anyon gauge.  

The statistical mechanics of the NACS particles 
can be studied in the low-density regime in terms of the cluster expansion 
of the grand partition function $\Xi$
\begin{equation} 
\Xi = \sum_{N=0}^\infty \nu^N\, {\rm Tr}\, e^{-\beta H_N}\,.
\label{par1}
\end{equation} 

The virial expansion (with the 
pressure expressed in powers of the density $\rho=\frac{N}{A}$) is given as  
\begin{equation} 
P= \rho k_B T\left[ 1+ B_2(T) \rho+ B_3(T) \rho^2 +\dots \right]\,, 
\end{equation} 
where $B_n(T)$ is the $n$-th virial coefficient, which can 
be expressed in terms of the first cluster coefficients $b_1,\cdots b_n$. 
The second virial coefficient $B_2(T)$ turns out to be \cite{Huang87}	  
\begin{equation} 
B_2(T) \,=\, -\frac{b_2}{b_1^2} 
\,=\, A\,\left(\frac{1}{2}-\frac{Z_2}{Z_1^2}\right)\,, 
\end{equation} 
where $A$ is the area and  
$Z_N={\rm Tr}\, e^{-\beta H_N}$ the $N$-particle partition function.
We assume that the NACS particles have 
equal masses and belong to the same isospin multiplet 
$\{|l,m>\}$ with $m=-l, \dots, l$. The quantity $Z_1={\rm Tr}\, e^{-\beta H_1}$ 
is then given by
\begin{equation} 
Z_1 = (2l+1) A/ \lambda^{2}_T \label{zone}\,. 
\end{equation} 

The computation of $Z_2= {\rm Tr}\, e^{-\beta H_2}$ is discussed in 
\cite{Lee95,Hagen96,Mancarella13}. It is convenient 
to separate the center-of-mass and relative coordinates: defining 
$Z = (z_1+z_2)/2$ and $z = z_1 -z_2$ one can write 
\begin{equation} 
H_2 = H_{\rm cm} + H_{\rm rel}=-\frac{1}{ 2\mu} \partial_Z \partial_{\bar Z} 
-\frac{1}{\mu}(\nabla_z\nabla_{\bar z} +\nabla_{\bar z}\nabla_z)\,\,\,, 
\end{equation} 
where $\mu\equiv M/2$ is the two-body reduced mass, 
$\nabla_{\bar z} = \partial_{\bar z}$ and 
$$
\nabla_z = \partial_z +\frac{\Omega}{z}\,\,\,. 
$$
$\Omega$ is a block-diagonal matrix given by  
$$
\Omega = \hat Q^a_1\hat Q^a_2 / (2\pi\kappa)=
\sum_{j=0}^{2l} \omega_j\otimes{I}_j\,\,\,, 
$$ 
with $\omega_j\equiv\frac{1}{4\pi\kappa} 
\left[j(j+1)-2l(l+1)\right]$. \; $Z_2$ can be then written as 
\begin{equation} 
Z_2= 2A  \lambda^{-2}_T Z_2^\prime \,, 
\label{hamrel:a}
\end{equation}
where $Z_2^\prime ={\rm Tr}_{\rm rel}\, e^{-\beta H_{\rm rel}}$. 
The similarity transformation 
$G(z,\bar z)  = \exp\left\{-\frac{\Omega}{2}\ln(z\bar z)\right\}$,  
acting as
\begin{eqnarray} 
H_{\rm rel} &\longrightarrow & 
H_{\rm rel}^\prime = G^{-1} H_{\rm 
rel} G,\nonumber\\ 
\Psi(z,\bar z) &\longrightarrow & 
\Psi^\prime(z,\bar z) =G^{-1} \Psi(z,\bar z)\,\,\,,\label{sim} 
\end{eqnarray} 
gives rise to an Hamiltonian $H_{\rm rel}^\prime$ 
manifestly Hermitian  
and leaves invariant $Z^\prime_2$.
The explicit expression for $H_{\rm rel}^\prime$ is
\begin{equation} 
H_{\rm rel}^\prime = -\frac{1}{\mu}(\nabla_z^\prime\nabla_{\bar z}^\prime 
+ \nabla_{\bar z}^\prime\nabla_z^\prime)\,\,\,,\label{trham}
\end{equation} 
where $\nabla_z^\prime = \partial_z + \Omega/2z$ and 
$\nabla_{\bar z}^\prime=\partial_{\bar z} -\Omega/2 {\bar z}$. 

By rewriting $H_{\rm rel}^\prime$ in polar coordinates and 
projecting it onto the subspace of total isospin $j,$ 
its correspondence with the Hamiltonian for (Abelian) 
anyons in the Coulomb gauge, having statistical 
parameter given by $\alpha_s= \omega_j$, becomes evident: 
\begin{equation} 
H_j^\prime = -\frac{1}{2\mu}\left[\frac{\partial^2}{\partial r^2}+ 
\frac{1}{r}\frac{\partial}{\partial r}+\frac{1}{r^2}\left(\frac{\partial} 
{\partial \theta}+i\omega_j\right)^2\right]\,. 
\end{equation} 
The same analysis discussed in Section \ref{abeliananyons} shows 
that the radial factor of the $j,j_z-$ component of the relative 
$(2l+1)^2-$vector wavefunction $\psi=e^{i n\theta} R_n(r)$ 
obeys the Bessel equation 
\begin{equation} 
\label{rsedfmh}
\frac{1}{M}\left[-\frac1r \frac{d}{dr}r\frac{d}{dr} + \frac{(n+\omega_j)^2}{r^2}\right] 
R^{j,j_z}_n(r) = ER^{j,j_z}_n(r) \equiv \frac{k^2}{M}R^{j,j_z}_n(r)\,,
\end{equation}   
whose general solution is  
\begin{equation} 
R^{j,j_z}_n(r) = A^{j,j_z}J_{|n+\omega_j|}(kr) + B^{j,j_z}J_{-|n+\omega_j|}(kr)\,\,\,. 
\end{equation} 

\section{Second Virial Coefficient: general soft-core case} 
\label{appendicesoftcore}
If one removes the hard-core boundary condition for the 
relative $(2l+1)^2$-component two-anyon wavefunction, and 
fixes an arbitrary external potential in order to regularize the spectrum, 
then the spectrum of each projected Hamiltonian operator $H_j'$ 
can be represented as the union of the spectra of $(2j+1)$ scalar 
Schr\"odinger operators, one for each $j_z$-component, 
endowed with its respective hard-core parameter 
$\varepsilon_{j,j_z}$. 
As discussed in Section \ref{nonabeliananyons}, 
one then ends up with a set of $(2l+1)^2$ (in principle independent) 
parameters $\varepsilon_{j,j_z}$, which are needed to fix the boundary behavior: 
\begin{equation} 
\left( 
\begin{array}{lllll} 
\varepsilon_{0,0} & \varepsilon_{1,1} & \varepsilon_{2,2} & \cdots & \varepsilon_{2l+1,2l+1}\\ 
\varepsilon_{1,-1} & \varepsilon_{1,0} & \varepsilon_{2,1} & \cdots & \varepsilon_{2l+1,2l}\\ 
\varepsilon_{2,-2} & \varepsilon_{2,-1} & \varepsilon_{2,0} & \cdots & \cdots \\ 
\cdots & \cdots & \cdots & \cdots & \cdots \\ 
\varepsilon_{2l+1,-2l-1} & \varepsilon_{2l+1,-2l} & \cdots & \cdots & \varepsilon_{2l+1,0} \\ 
\end{array} 
\right)\,. 
\label{matrix}
\end{equation} 
For the general soft-core NACS gas, one has the following expression for the 
second virial coefficient \cite{Mancarella13}:
\begin{equation} 
B_2^{s.c.}\left( \kappa,l,T \right)=
\frac{1}{(2l+1)^2}\sum_{j=0}^{2l}\sum_{j_z=-j}^{j}\left[\frac{1+(-1)^{j+2l}}{2} B^B_2(\omega_j,T,\varepsilon_{j,j_z}) + 
\frac{1-(-1)^{j+2l}}{2}B^F_2(\omega_j,T,\varepsilon_{j,j_z})\right]\,\,\,,
\label{generalformulasoftcorecase} \end{equation} 
where $B^B_2(\omega_j,T,\varepsilon_{j,j_z})$ 
is the soft-core expression entering Eq. (\ref{explicitintegralform}):
\begin{equation} 
B^B_2(\omega_j,T,\varepsilon_{j,j_z})=B_2^{h.c.}(\delta_j,T)- 2 \lambda_T^2 
\left\{ 
e^{\varepsilon_{j,j_z}} \theta(-\sigma)  
+\frac{\delta_j\sigma}{\pi} \left(\sin{\pi \delta_j} \right)
\int_0^\infty \frac{dt e^{-\varepsilon_{j,j_z}\, t} t^{\vert\delta_j\vert-1}}{1+2\sigma(\cos{\pi\delta_j})\;t^{|\delta_j|}+t^{2|\delta_j|}} \right\},
\end{equation}
with $\delta_j\equiv (\omega_j+1) \, mod \, 2-1$, and 
$B^F_2(\omega_j,T,\varepsilon_{j,j_z})$ is the previous expression 
evaluated for $\omega_j \rightarrow \omega_j+1$:   
\begin{equation} 
B^F_2(\omega_j,T,\varepsilon_{j,j_z})=B_2^{h.c.}(\Gamma_j,T)- 2 \lambda_T^2 
\left\{ 
e^{\varepsilon_{j,j_z}} \theta(-\sigma)  
+\frac{\Gamma_j \,\sigma}{\pi} \left(\sin{\pi \Gamma_j} \right)
\int_0^\infty \frac{dt e^{-\varepsilon_{j,j_z}\, t} t^{|\Gamma_j|-1}}{1+2\sigma(\cos{\pi\Gamma_j})\;t^{|\Gamma_j|}+t^{2|\Gamma_j|}} \right\},
\end{equation}
with $\Gamma_j\equiv \omega_j \, mod \, 2\,-1$. 

To perform explicit computations, 
we may consider the simple case 
in which an isotropy for the hard-core parameter 
is assumed within each shell with assigned isospin quantum number $l$.  
In other words, $\varepsilon_{j,j_z}\equiv\varepsilon_j$ and the matrix 
(\ref{matrix}) then reads
\begin{equation} 
\varepsilon_{j,j_z}\equiv\left( 
\begin{array}{llll} 
\varepsilon_{0} & \varepsilon_{1} & \cdots & \varepsilon_{2l+1}\\ 
\varepsilon_{1} & \varepsilon_{1} & \cdots & \varepsilon_{2l+1}\\ 
\cdots & \cdots & \cdots & \cdots \\ 
\varepsilon_{2l+1} & \varepsilon_{2l+1} & \cdots  & \varepsilon_{2l+1} \\ 
\end{array} 
\right)\,. 
\label{matrix_is}
\end{equation} 
When all of the elements of the matrix (\ref{matrix_is}) are equal, we will 
use the notation $\varepsilon_{j,j_z} \equiv \varepsilon$. 
In such a completely isotropic case, 
Eq. (\ref{generalformulasoftcorecase}) takes the 
simplified form
\begin{equation} 
\label{simplifiedsoftcore}
B_2^{s.c.}\left(\kappa,l,T\right)=
\frac{1}{(2l+1)^2}\sum_{j=0}^{2l}(2j+1)\,B^B_2(\nu_j,T,\varepsilon)\,\,\,,
\end{equation}
where 
\begin{equation}
\label{modifiedparameter}
\nu_j\equiv\left(\omega_j-\frac{1+(-1)^{j+2l}}{2}\right) \, mod \, 2-1\,.\\
\end{equation}
 
For $l=1/2$, i.e. the lowest possible value for non-Abelian anyons, 
the assumption of isotropy ($\varepsilon_{0,0}=\varepsilon_0$ 
and $\varepsilon_{1,m}=\varepsilon_1$ with $m=1,0,-1$) yields: 
\begin{equation} 
B_2^{s.c.}\left(\kappa,l=\frac{1}{2},T\right)=\,\frac{3}{4}B^B_2(\omega_1,T,\varepsilon_{1}) +\frac{1}{4}B^F_2(\omega_0,T,\varepsilon_{0})\,. 
\label{isotropictwodimensional} 
\end{equation} 
As an example, let us consider the case $l=1/2, 4 \pi \kappa=3$:    
\begin{equation} 
B_2^{s.c.}\left(k=3,l=\frac{1}{2},T\right)=\,\frac{3}{4}B^B_2
\left(\alpha=\frac{1}{6},T,\varepsilon_{1}\right) + 
\frac{1}{4}B^F_2\left(\alpha=-\frac{1}{2},T,\varepsilon_{0}\right)\,\,\,. 
\label{esempio} 
\end{equation} 
If the four parameters of the whole matrix are taken to be identical 
$\varepsilon_0 =\varepsilon_1\equiv\varepsilon$ 
["complete isotropy" of the parameter matrix (\ref{matrix})], 
the virial coefficient reduces to  
\begin{equation} 
B_2^{(s.c.)}\left(k=3,l=\frac{1}{2},T,\sigma=1\right)=-\frac{\lambda_T^2}{24}\left\{1+\frac{4}{\pi}\int_0^\infty dt\; e^{-\varepsilon t} \left(\frac{6 \; t^{-1/2}}{1+t}+\frac{t^{-5/6}}{1+\sqrt{3}\,t^{1/6}+t^{1/3}}\right) \right\}\,\,\,. 
\label{fiboisotro} 
\end{equation} 
The depletion of $B_2$ in Eq. (\ref{fiboisotro}), 
with respect to its hard-core value $-\frac{1}{24}\lambda_T^2$, 
arises from the anyonic collisions allowed by the soft-core conditions.


\begin{thebibliography}{XX}
\bibitem{note} We are indebted to Y. Stein for pointing out this point.
\end{thebibliography}
\end{document}